\documentclass[aps,pre,onecolumn,showkey,square,numbers,amssymb,amsmath,nobibnotes,footinbib]{revtex4}
\usepackage[utf8]{inputenc}
\usepackage{bm}
\usepackage{amsmath}
\usepackage{mathtools}
\usepackage{epsfig}
\usepackage{bbold}
\usepackage{eucal}
\usepackage{graphicx}
\usepackage{float}
\usepackage[utf8]{inputenc}
\usepackage{hyperref}
\usepackage[font=small]{caption}
\usepackage{subcaption}
\usepackage{color}
\usepackage{comment}
\newfloat{figure}{htbp}{figs}
\newcommand{\bOmega}{{\bm \Omega}}
\newcommand{\I}{{\mathrm{i}}}

\begin{document}
\title{Rationalizing systematic discrepancies between election outcomes and opinion polls}
%~

\author{Luca Gamberi}
\author{Pierpaolo Vivo}
\author{Yanik-Pascal F\"{o}rster}
\author{Evan Tzanis}
\author{Alessia Annibale}

\affiliation{Quantitative and Digital Law Lab, Department of Mathematics, King's College London,
	Strand, WC2R 2LS, London (United Kingdom)}

\homepage[]{Quantlaw.co.uk}

\begin{abstract}
The Bradley effect concerns the discrepancy between opinion polls and actual election outcomes that emerges when candidates do not exhibit ideological, sexual or racial mainstream features. This effect was first observed during the 1982 election for the Governor of California that resulted in a significant loss for the black Democratic candidate, Tom Bradley, despite him being ahead in polls. It has been argued that poll respondents may tend to mask their true political preference %for non-canonical candidates in
in favour of what is generally considered more socially acceptable. We propose an exactly solvable statistical mechanical model, which allows for a quantitative exploration of this phenomenon. The model includes three main ingredients: (i) the tendency of individuals to align their real preference to the declared (public) opinions of others, (ii) a term accounting for an individual \emph{integrity} factor, which induces voters to be consistent between their public and private opinions, and (iii) a \emph{self-monitoring} component, which tunes the strength by which an individual is affected by and wishes to publicly conform to social norms. We observe the emergence of order in a wide range of parameters of the model and discuss the effect of noise on the correlation between hidden and public preference. Finally, we present an application of this model to the prediction of election outcomes.
\end{abstract}

\maketitle

%\noindent{\it Keywords\/}: 

\section{Introduction}
In the 1982 election for the Governor of California, the long-time Democratic mayor of Los Angeles, Tom Bradley, narrowly lost the race against the Republican candidate George Deukmejian in spite of leading in the polls in the final days before the election \cite{WP82}. Since then, the expression ``\emph{Bradley effect}" has been used by political commentators to refer to the -- somewhat controversial -- phenomenon by which significant discrepancies are observed between voter opinion polls and actual election outcomes where a white and a non-white candidate run against each other \cite{St08}. The Bradley effect is regularly mentioned every time pollsters are caught by surprise by the true outcome of binary elections, which they had predicted would lead to a comfortable majority for the black candidate -- when the margin turns out to be significantly thinner in reality, or is reversed altogether (see e.g. the 1983 race for Mayor of Chicago, the 1988 Democratic primary race in Wisconsin for President of the United States, or the 1989 race for Mayor of New York City).

Lately, the Bradley effect has been related to a wider range of scenarios in political polls where there is a general behavioral tendency to conceal the true opinion when that could lead to criticism. This effectively represents the electoral incarnation of the so-called \emph{social desirability bias} (SDB) \cite{Grim10, BLM54,LL96}. The related concept of \emph{preference falsification} \cite{Kura95} has been portrayed as one of the main drivers of paradoxical behaviors whereby individuals openly support norms or political systems they privately loathe \cite{Kura95,N-N74,JY16}, a tendency that can be demonstrated in behavioral and numerical experiments too \cite{MBKS80,CWM05,WKM09}.

SDB therefore represents one of the main challenges that pollsters face, acting as one of the main drivers of uncertainty in predictions together with the so-called \emph{hidden vote}, which occurs when a great number of predicted abstainers turn up at polling stations, making historical polls data effectively useless \cite{ZSCCM21,U-S06}. 

Negative effects deriving from the SDB span far beyond the remit of political systems and elections. SDB is well-known as a potentially dangerous source of distortion in surveys and self-reports as well, whereby certain individual traits, beliefs or even efforts are arbitrarily inflated or deflated by each respondent in an unconscious attempt to please the interviewer by portraying a better version of one self, compromising the validity of the surveys and reports. The effect was observed to happen, for instance, in dietary reporting \cite{HCPOO95}, workouts tracking \cite{AMEMCF05} or even in self-reporting housekeeping work division within a family \cite{PT98,Kamo00}.

Several methods have been devised to mitigate the effects of such distortions such as improving anonymity and confidentiality, randomizing and/or grouping responses, and deflecting the attention from the respondents' behavior to the close circles of their acquaintances \cite{Grim10,Nede85}, and, although these polling techniques can dampen the effects of SDB over the poll outcome, they are not able to remove them completely \cite{ZSCCM21}.

While the socio-physics literature has produced a wealth of rather sophisticated models to investigate \emph{opinion dynamics}, namely how one agent's beliefs are updated based on what \emph{others} are thinking, more limited effort has been devoted so far to the quantitative understanding of the discrepancy between one's \emph{own} public and private universe. In this paper, we present an exactly solvable statistical mechanical model, in which voters need to cast a (secret) preference for two candidates and -- at the same time -- a publicly available declaration of support for either one of them (allowing for the possibility of refusing to express a public preference).

The dynamics resulting from human interactions whereby one's own opinions and behaviors are shaped by the presence (physical or virtual) of others, has been investigated at length in the psychology and sociology literature \cite{Lat81}. Structural models, which are popular theories in sociology for explaining naturally occurring homogeneity in society, display the same characteristics adopted in complex network theories of social dynamics \cite{Fried84,Fried99,Fried03,FJ99}. In \cite{Mosc80}, Moscovici argues that this emergence of consensus in human behaviors may be ascribed to two elements of the human psychology: the tendency of an individual to \textit{comply} by not necessarily making the group's belief their own and, complementary, the tendency to be entirely \textit{converted} by the group's opinion.

In applications of statistical mechanics, an opinion is  described by a dynamical state variable that can be continuous or discrete, which represents the individual inclination, at a given time, towards a certain issue or idea to which she/he is exposed. The system of individuals (agents) interact via ``influence fields" that often percolate through a network structure \cite{BLT15,SLST16,IKKB09}. Opinions dynamics models of this sort were proposed by looking at several parameters e.g. peer-pressure \cite{SGSR19}, memory and learning \cite{JS-W18,AO11}, polarization and extremism \cite{SSVL20,BL-SSS20}, the effect of bounded confidence in continuous systems \cite{Lore07}, to cite a few. 

One of the most successful models of opinion dynamics is the so-called \emph{voter model}, which is based, in its simplest form, on reasonable and often analytically tractable assumptions. In voter models, agents are capable of expressing a binary preference and can change their opinion by taking up that of one of their neighbors. In particular, an agent's preference changes according to a stochastic rule: whoever is selected (at random) adopts one of its neighbor's opinion with a given probability, regardless of its own previous opinion. For its simplicity and richness, the voter model has been extensively studied and several modifications of the original model have been proposed.  
For reviews, see \cite{SLST16,Ligg99,CFL09}.

Closely related to this study is the \emph{concealed voter model} (CVM), introduced in \cite{GOG18}, 
which posits that the opinion an individual publicly expresses may well be different from its own inner beliefs. Each agent $i$ is associated with an opinion doublet, containing the concealed and public opinion. Cross-agent interaction takes place only through their public opinion via imitation (similarly to the classic voter model). This dynamical rule competes with \emph{introspection}, a mechanism that allows one's own public opinion to relax into the concealed state and vice-versa. Ref. \cite{GTGSO19} studied the effect of the level of hypocrisy on this model. A variation of the CVM describing opinion dynamics following irrational (or rational) behaviors was presented in \cite{JMNS-W18}.

The majority of opinion dynamics models, which are realised either following a game-theoretic approach or as variations of the dynamical Ising model, share the common objective of studying the time scale at which consensus (or other steady states) is reached \cite{CFL09}. Another strand of literature has instead focused on \emph{equilibrium} statistical mechanics models for the emergence of collective behaviour and opinion spreading in society \cite{Galam08,CG07,CGM08,B-NKR03,DGM02}. 

Our contribution to the existing literature is to explicitly model the interactions of a spin system in equilibrium taking place on a bipartite network, similarly to \cite{GOG18}.  In particular, we include interaction terms that account for the different ways the publicly available (expressed) opinion affects one's own inner belief. Moreover, our model prescribes a ``neutral'' state for the expressed opinion to reflect unsampled voters in the polls.

In section \ref{sec:model} we introduce the bipartite spin system, define its equilibrium Hamiltonian and compute the free energy. In section \ref{sec:eq_analisys}, we study the properties of the system in equilibrium and show that for a range of values for the physical parameters one can observe vote polarization, and in particular, a Bradley effect-like scenario in which declared and true preferences have opposite sign. In \ref{sec:self_align} we present the effect of introducing a social temperature which, for some values of the couplings, creates a disagreement region between election results and self-consistency. In section \ref{sec:beyond_avg} we compute the exact finite $N$ distribution for the magnetization in both layers which are then employed in \ref{sec:ML_ms} for finding the maximum likelihood estimation of the true vote, given a set of historical election data. Finally, we present our results and discuss the applicability of the model in section  \ref{sec:conclusion}.

\section{Model outline}\label{sec:model}
We consider a system of $N$ interacting voters, each making a binary choice $\sigma_i\in\{+1,-1\}$ expressed in the secrecy of the polling station. The vote outcome is represented by the set $\bm\sigma=(\sigma_1,\ldots,\sigma_N)$. The same population is the target of opinion polls, in which they may declare a vote $\tau_i\in \{+1,0,-1\}$, with $\tau_i=0$ corresponding to voter $i$ refusing to declare a vote or not being polled which ensures that the poll outcome set $\bm\tau=(\tau_1,\ldots,\tau_N)$ contains a random variable for each voter.

\begin{figure}[ht]
    \centering
    {\includegraphics[width=0.5\textwidth]{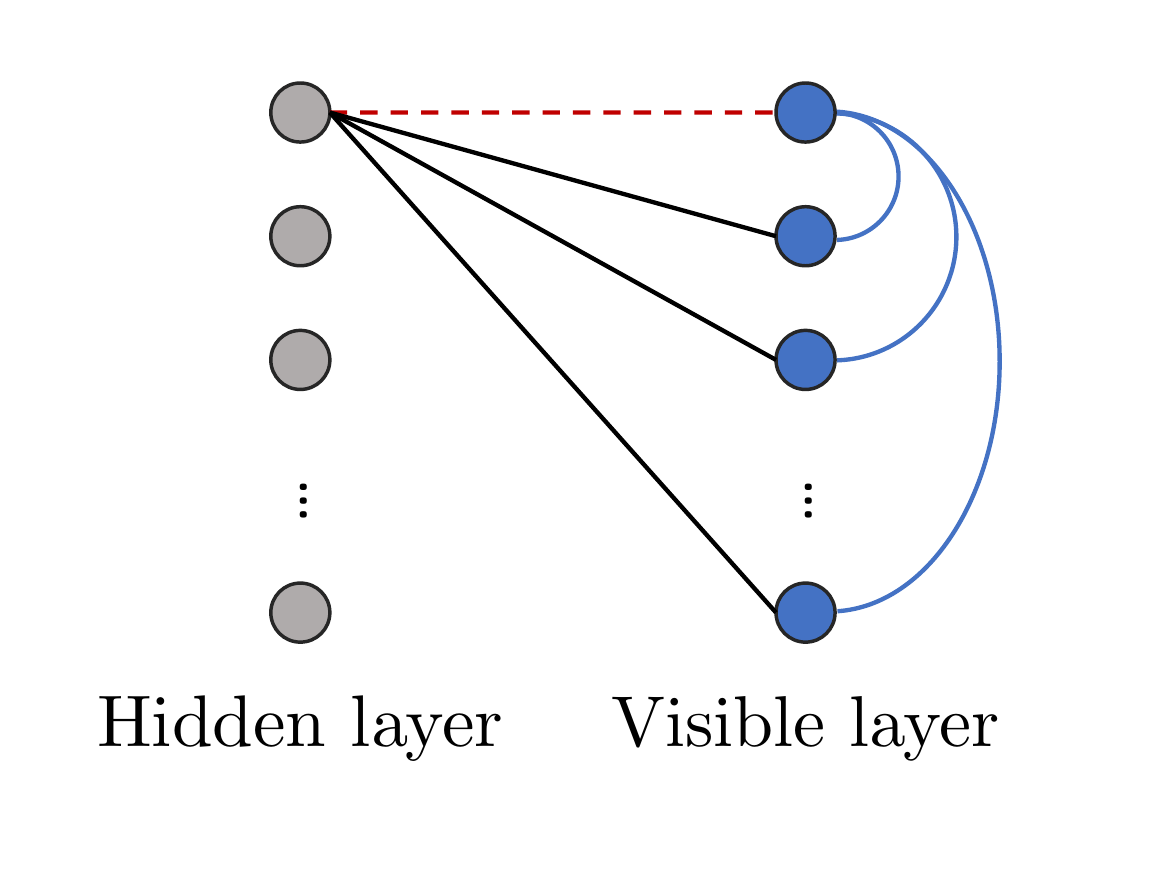}}
\caption{Schematic representation of the $N$ voter system. The  dashed red line represent the channel through which the integrity interaction occurs, the black and blue lines represent the conversion and compliance interaction channels.}
\label{fig:bipartite}
\end{figure}

We model this system via a bipartite graph, in which one layer, namely the hidden layer, represents the true votes $\bm\sigma$, and the visible layer consists of the set of publicly expressed opinion $\bm\tau$, as shown in figure \ref{fig:bipartite}. Since we assume that the voting process is anonymous and that the outcome is revealed ex-post, no direct interaction occurs within the set $\bm\sigma$. However, since the set $\bm\tau$ is visible to everyone, each declared vote will influence both the other public declarations and the internal (true) preferences.

Our model incorporates three features that have been identified as important in influencing individuals' behaviour or beliefs when exposed to others' opinions \cite{Mosc80}:
\begin{itemize}
\item The \textit{conversion strength} being the tendency of voters to conform their real belief to the publicly available preferences of others (or anti-conform);
\item The level of \emph{social integrity} (or \emph{deception}) each voter is capable of displaying, i.e. the tendency to align (or anti-align) their declared preference with their real one;
\item The \textit{compliance strength}  which controls  one's own desire to comply to (or to challenge) the ``wisdom of the crowd", i.e. to align (or anti-align) their declared preference to the others', without necessarily changing their true preference.
\end{itemize} 

\noindent
These competing mechanisms can be modelled by the following Hamiltonian
\begin{equation}
H=-\frac{J}{N}\sum_{i\neq j}^N \sigma_i\tau_j - \omega\sum_{i=1}^N  \sigma_i\tau_i - \frac{\phi}{N}\sum_{i\neq j}^N \tau_i\tau_j \ ,
\label{eq:interact_H}
\end{equation}
with $J$ the conversion strength, $\omega$ the social integrity, and $\phi$ the compliance strength.

The system of spins is assumed to be in equilibrium, at inverse \emph{social temperature} $\beta$ which denotes the degree of noise, i.e. of irrational, non-deterministic behaviour of voters, studied in \cite{LLW05}. Equilibrium is reached when enough time is passed from an opinion poll and the voters have had time to ``thermalise" to form their true opinion, which is then cast into vote.

For the sake of simplicity we consider the conversion and compliance strengths and integrity to be constant throughout the system. The joint probability distribution for the equilibrium system of spins $\bm\tau$ and $\bm\sigma$ is given by
\begin{equation}
P(\bm\sigma,\bm\tau)=\frac{1}{Z_N}\exp\left\{\frac{\beta J}{N}\sum_{i\neq j}^N\sigma_i\tau_j+\beta \omega\sum_{i=1}^N \sigma_i\tau_i+\frac{\beta\phi}{N} \sum_{i\neq j}^N \tau_i\tau_j\right\} \ ,
\label{eq:joint_p}
\end{equation}
with $Z_N$ being the partition function, for a system size $N$, defined as follows
\begin{equation}
Z_N
= \sum_{\bm\sigma,\bm\tau}\exp\left\{\frac{\beta J}{N}\sum_{i\neq j}^N\sigma_i\tau_j+\beta \omega\sum_{i=1}^N \sigma_i\tau_i+\frac{\beta\phi}{N} \sum_{i\neq j}^N \tau_i\tau_j\right\}\ .
\label{eq:joint_p_approx}
\end{equation}

In the large $N$ limit, we can compute exactly the free energy density $f(\beta,J,\omega,\phi)$,
which gives access to the physical properties of the system
at equilibrium, as follows
\begin{align}
f(\beta,J,\omega,\phi) &= - \lim_{N\rightarrow \infty}\frac{1}{\beta N}\log Z_N\nonumber\\
&=- \lim_{N\rightarrow \infty}\frac{1}{\beta N}\log 
\int \frac{dm_\tau dm_\sigma}{N^2} e^{\beta N (Jm_\tau m_\sigma + \phi m_\tau^2)}\ \sum_{\bm\sigma,\bm\tau} e^{\beta\omega\sum_{i} \sigma_i\tau_i} \delta(Nm_\sigma - \sum_i \sigma_i)\delta(Nm_\tau - \sum_i \tau_i)\nonumber\\ 
&= - \lim_{N\rightarrow \infty}\frac{1}{\beta N}\log \int \frac{dm_\sigma d\xi dm_\tau d\zeta}{(2\pi/N)^2} e^{\beta N (Jm_\tau m_\sigma + \phi m_\tau^2)}\sum_{\bm\sigma,\bm\tau} e^{\beta\omega\sum_{i} \sigma_i\tau_i -\I\xi(Nm_\sigma -\sum_i \sigma_i) -\I\zeta(Nm_\tau -\sum_i \tau_i)}\nonumber\\
&=- \lim_{N\rightarrow \infty}\frac{1}{\beta N}\log \int \frac{dm_\sigma d\xi dm_\tau d\zeta}{(2\pi/N)^2} e^{Ng(m_\sigma,m_\tau,\xi,\zeta)} \ ,
\label{eq:saddle_point_expansion_Z}
\end{align}
where we have used the Fourier representation of the delta functions with auxiliary variable $\xi$ and $\zeta$ and performed the summations over $\bm\sigma$ and $\bm\tau$ to obtain
\begin{equation}
    g(m_\sigma,m_\tau,\xi,\zeta) = \beta Jm_\tau m_\sigma +\beta \phi m_\tau^2 -\I\xi m_\sigma - \I\zeta m_\tau +\log[e^{\I\xi} (1+2 \cosh(\beta \omega+ \I\zeta)) + e^{-\I\xi} (1+2 \cosh(-\beta \omega+ \I\zeta))]\ .
\end{equation}
Evaluating the integral in \eqref{eq:saddle_point_expansion_Z} via steepest descent we find
\begin{align}
    f(\beta,J,\omega,\phi)
    & = - \frac{1}{\beta}\lim_{N\rightarrow \infty}  g(m_\sigma^\star,m_\tau^\star,\xi^\star,\zeta^\star)\nonumber\\ & = J m_\sigma^\star m_\tau^\star + \phi m_\tau^{\star2} -\frac{1}{\beta} \mathrm{log}\left[ e^{\beta J m_\tau^\star}\left(1+2\cosh\left(\beta \omega+\beta J m_\sigma^\star +2\beta \phi m_\tau^\star\right)\right)+\right.\nonumber\\
    &\left.+e^{-\beta J m_\tau^\star}\left(1+2\cosh\left(-\beta \omega+\beta J m_\sigma^\star +2\beta \phi m_\tau^\star\right)\right)\right]\ ,
    \label{eq:free_energy}
\end{align}
where we have expressed the conjugate variable $\xi^\star$ and $\zeta^\star$ in terms of the magnetizations using the stationary conditions $\mathrm{i}\xi^\star=\beta J m_\tau^\star$ and $\mathrm{i}\zeta^\star =\beta J m_\sigma^\star +2\beta \phi m_\tau^\star$. The stationary conditions over the magnetizations lead to the coupled fixed point equations
\begin{align}
\begin{dcases}
m_\sigma^\star =\frac{e^{\beta J m_\tau^\star}\left(1+2\cosh\left(\beta \omega+\beta J m_\sigma^\star +2\beta \phi m_\tau^\star\right)\right)-e^{-\beta J m_\tau^\star}\left(1+2\cosh\left(-\beta \omega+\beta J m_\sigma^\star +2\beta \phi m_\tau^\star\right)\right)}{e^{\beta J m_\tau^\star}\left(1+2\cosh\left(\beta \omega+\beta J m_\sigma^\star +2\beta \phi m_\tau^\star\right)\right)+e^{-\beta J m_\tau^\star}\left(1+2\cosh\left(-\beta \omega+\beta J m_\sigma^\star +2\beta \phi m_\tau^\star\right)\right)}\\
m_\tau^\star =2\frac{e^{\beta J m_\tau^\star}\sinh\left(\beta \omega+\beta J m_\sigma^\star +2\beta \phi m_\tau^\star\right)+e^{-\beta J m_\tau^\star}\sinh\left(-\beta \omega+\beta J m_\sigma^\star +2\beta \phi m_\tau^\star\right)}{e^{\beta J m_\tau^\star}\left(1+2\cosh\left(\beta \omega+\beta J m_\sigma^\star +2\beta \phi m_\tau^\star\right)\right)+e^{-\beta J m_\tau^\star}\left(1+2\cosh\left(-\beta \omega+\beta J m_\sigma^\star +2\beta \phi m_\tau^\star\right)\right)}
\end{dcases}
\ .
\label{eq:magn_state}
\end{align}
Since we are interested in the maxima of $g$ (corresponding to the minima of the free energy density), in the next sections we will determine its stationary values and  analyze their stability properties.
This will give us information on the voting (and declaration) polarization. We will also analyze the voter's ``self-alignment'', i.e. the average consistency between one's own declared and true vote.

\section{Equilibrium analysis}\label{sec:eq_analisys}
The system \eqref{eq:magn_state} admits a paramagnetic solution at  $(m_\sigma^\star,m_\tau^\star)=(0,0)$, for all values of $\beta$, although this solution is expected to become unstable at low temperatures. Given the symmetry of the hyperbolic functions, it is easy to check that the equations in \eqref{eq:magn_state} are odd symmetric and thus we expect solutions in the form $\pm (m_\sigma^\star,m_\tau^\star)$.

\begin{figure}
    \centering
    \includegraphics[width=0.5\textwidth]{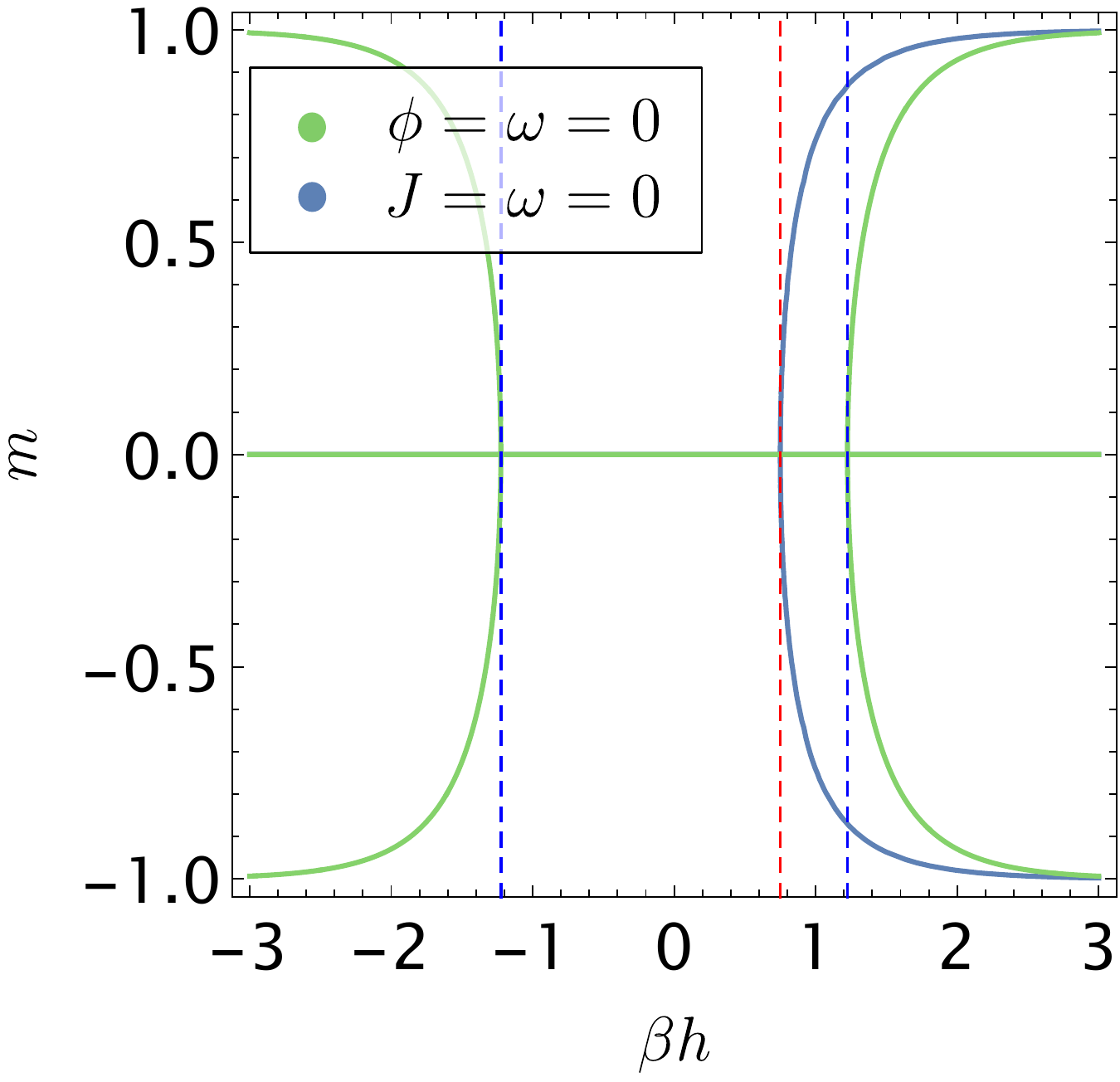}
\caption{Solutions for the magnetizations in the limiting cases. In particular, the green curve corresponds to magnetization $m_\sigma$ for case $\phi=\omega=0$ (with $h=J$) and the blue curve to the magnetization $m_\tau$ for $J=\omega=0$ (with $h=\phi$). The vertical dashed lines indicate the critical values for the effective couplings $\beta\phi=3/4$, in red, and $\beta J =\pm\sqrt{3/2}$, in light blue.}
\label{fig:m_limit}
\end{figure}

Let us first check the limiting case in which the interaction only takes place on the visible layer, i.e. $J=\omega=0$. In this limit, the Hamiltonian reduces to
\begin{equation}
H=-\frac{\phi}{N}\sum_{i\neq j}^N \tau_i\tau_j \ .
\end{equation}
Were $\tau=\pm 1$ this would be an Ising ferromagnet with inverse critical temperature $\beta=1/\phi$. The Ising ferromagnet with the extra degree of freedom $\tau=0$ corresponds to a mean-field Blume-Capel model \cite{cap66}. From the saddle-point equation for the magnetization $m_\tau^\star$, setting $J=\omega=0$, one gets
\begin{equation}
    m_\tau^\star = 2\frac{\sinh(2\beta\phi m_\tau^\star )}{1 + 2 \cosh(2\beta\phi m_\tau^\star)}\ .
    \label{eq:m_limit_phi}
\end{equation}
Macroscopic order appears for couplings over the critical value $\beta\phi \geq 3/4$, as shown in figure \ref{fig:m_limit}.

Another limiting case is obtained by setting $\phi=\omega=0$, which corresponds to the bipartite system
\begin{equation}
H=-\frac{J}{N}\sum_{i\neq j} \sigma_i\tau_j \ ,
\label{eq:H_bip}
\end{equation}
where $J$ is the strength of the bipartite (ferromagnetic) interactions. In this case, the system \eqref{eq:magn_state} reduces to
\begin{align}
\begin{dcases}
 m_\tau^\star = 2\frac{\sinh(\beta J m_\sigma^\star )}{1 + 2 \cosh(\beta J m_\sigma^\star)}\\
 m_\sigma^\star = \tanh\left(2 \beta J  \frac{\sinh(\beta J m_\sigma^\star )}{1 + 2 \cosh(\beta J m_\sigma^\star)}\right)
\end{dcases}
\ .
\label{eq:m_limit_J}
\end{align}
Comparing the two equations in \ref{eq:m_limit_J}, one sees that $m_\sigma^\star$ is fully determined by $m_\tau^\star$ as the former can be expressed in terms of the other to give $m_\sigma^\star=\tanh\left(\beta J m_\tau^\star\right)$. The critical value for the effective coupling is found expanding the second equation in \ref{eq:m_limit_J} around $m_\sigma^\star=0$, which leads to a symmetric solution $(\beta J)^2=3/2$. The bifurcation diagram for $m_\sigma$ is shown in figure \ref{fig:m_limit}.

In both these limiting cases, the free energy is determined by one order parameter, the magnetization $m_\tau$ in the first, and $m_\sigma$ in the second.  Due to its symmetry, odd derivatives of $f$ are zero so its expansion reads
\begin{equation}
    f(\beta)=f_0 + \alpha m^2 + \frac{1}{2}\gamma m^4 + \ldots \ .
    \label{eq:landau_expansion}
\end{equation}
When $\gamma>0$, which ensures that minima are found for finite values of the magnetization, the stability is simply determined by the sign of $\alpha$:
\begin{itemize}
    \item for $J=\omega=0$, it is easily checked that when $0<\beta\phi<3/4$, $m_\tau=0$ corresponds to a minimum. For values of the coupling $\beta\phi<0$ the system does not bifurcate as the right hand side of equation \eqref{eq:m_limit_phi} is odd symmetric.
    The analysis of the second derivative of $f$ in this regime shows $m_\tau=0$ to be unstable and no other extremes are found for finite values of $m_\tau$.
    
    \item for $\phi=\omega=0$, the expansion in terms of $m_\sigma$ shows that when $-\sqrt{3/2}<\beta J<\sqrt{3/2}$, $m_\sigma=0$ corresponds to a minimum of $f$ in the interval. 
    
\end{itemize}

\subsection{Stationary states: One-layer polarization}
We now turn to analyzing the model in the full parameter space. First off, it is worth checking whether solutions corresponding to polarization of a single layer exist i.e. solutions of the form $(m_\sigma^\star,0)$ or $(0,m_\tau^\star)$. The former case is easily checked: by setting $m_\tau^\star=0$, the second equation of the system \eqref{eq:magn_state} becomes  $\sinh\left(\beta \omega+\beta J m_\sigma^\star \right)=\sinh\left(\beta \omega-\beta J m_\sigma^\star \right)$, which is solved by $m_\sigma^\star = 0$, consistently to what we previously observed for $\phi=\omega=0$.
Intuitively, this is consistent with the case in which the general public opinion does not offer a clear direction ($m_\tau^\star=0$) and thus true votes are also mixed and no consensus can emerge from this picture. 

\begin{figure}
    \centering
    \includegraphics[width=0.45\textwidth]{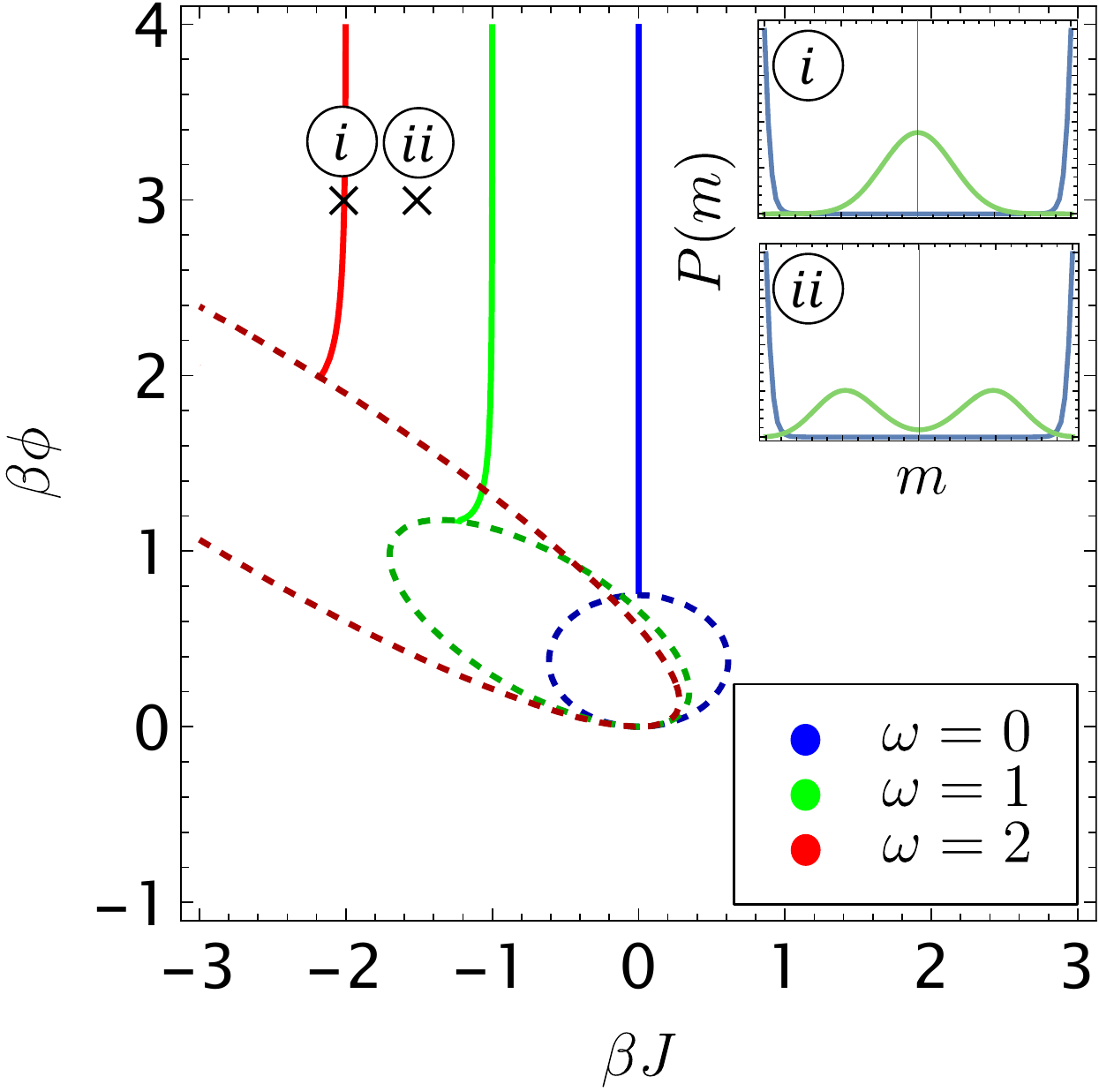}
\caption{Stability region of the solution $(0,m_\tau)$ in the parameter space $\beta J-\beta \phi$ at $\omega=\{0,1,2\}$. Within the region delimited by the dashed lines, the stable solution is paramagnetic. In the outer region, a non-trivial stable solution is found along the solid lines (corresponding each to a different value of $\omega$) and, elsewhere, the solution $(0,m_\tau)$ is unstable. Note that the case $\omega=0$, blue line, falls back on the result found in the limiting case $\omega=J=0$. The panels to the right show the distributions $p(m_\sigma)$ and $p(m_\tau)$ at $\beta\phi = 3$, $\beta\omega=2$ with ($i$) $\beta J \simeq -2.01$, corresponding to the stable one-layer polarization and, with ($ii$) $\beta J = -1.5$ that shows polarization in both layers.}
\label{fig:ms0}
\end{figure}

The latter case is obtained by setting $m_\sigma^\star=0$ and results in the system 
\begin{align}
\begin{dcases}
0 = e^{\beta J m_\tau^\star}\left(1+2\cosh\left(\beta \omega+2\beta \phi m_\tau^\star\right)\right)-e^{-\beta J m_\tau^\star}\left(1+2\cosh\left(-\beta \omega+2\beta \phi m_\tau^\star\right)\right)\\
m_\tau^\star =2\frac{e^{\beta J m_\tau^\star}\sinh\left(\beta \omega+2\beta \phi m_\tau^\star\right)+e^{-\beta J m_\tau^\star}\sinh\left(-\beta \omega+2\beta \phi m_\tau^\star\right)}{e^{\beta J m_\tau^\star}\left(1+2\cosh\left(\beta \omega +2\beta \phi m_\tau^\star\right)\right)+e^{-\beta J m_\tau^\star}\left(1+2\cosh\left(-\beta \omega +2\beta \phi m_\tau^\star\right)\right)}
\end{dcases}
\label{eq:magn_ms0}
\ ,
\end{align}
which can be solved numerically. Stable solutions are plotted, for different values of $\omega$, in Figure \ref{fig:ms0} (full lines).
The stability is checked by expanding (to the first order) the equations in system \eqref{eq:magn_state} around $m_\sigma=0$ and $m_\tau=m_\tau^*$ where $m_\tau^*$ is obtained numerically solving \eqref{eq:magn_ms0} in implicit form.

We can use the expansion of the free energy \eqref{eq:landau_expansion} around $m_\tau=0$, to look at the corresponding critical curve in the parameter space. First, we inspect the case $\omega=0$, where the coefficient $\alpha$ become zero for
    \begin{equation}
    \left(2\sqrt{\frac{2}{3}}\beta J\right)^2 + \left(\frac{8}{3}\beta\phi - 1\right)^2=1\ ,
    \end{equation}
    corresponding to an ellipse of critical points in the coupling space. Expanding up to the fourth order and studying the sign of $\gamma$, one gets that the solution $(0,0)$ is stable in the inner region of the ellipse. This region is delimited by the dashed lines in figure \ref{fig:ms0}.
    Switching on the coupling $\omega$ leads to a similar scenario in which the critical ellipse is given by the more complicated expression
    \begin{equation}
        \beta\phi - \frac{2((\beta J)^2+(2\beta\phi)^2)\cosh(\beta\omega)+\beta J (\beta J + 8 \beta\phi \sinh(\beta\omega))}{2+4\cosh(\beta\omega)}=0\ .
        \label{eq:elipse}
    \end{equation}
The insets in figure \ref{fig:ms0} show the distributions $p(m_\sigma)$ and $p(m_\tau)$, respectively, at $\beta\phi = 3$, $\beta\omega=2$ and two different values of $\beta J$, namely (a) $\beta J \simeq -2.01$, corresponding to the stable solution $(0,m_\tau)$ and, (b) $\beta J = -1.5$, where the solution $(0,m_\tau)$ is expected to be unstable. In this second case, both distributions are bimodal, suggesting
polarization in both layers.

The existence of a one-layer stable polarization suggests that in the absence of a clear public opinion, true votes do not polarize, while the opposite is not necessarily realized. In fact, for a combination of the system parameters $(\beta, J,\omega,\phi)$ we observe a non-zero solution for $m_\tau^\star$. This happens, for instance, for positive compliance and integrity but negative values for the conversion strength: a drive to display a socially acceptable opinion, mixed with moral probity, wrestles with the inability to convert one's inner belief to conform with the mass, leading to an apparent polarization of the consensus which does not materialize in the ballot box. Interestingly, in this case the declared preference could polarize towards either of the two candidates/choices, as one would expect due to the symmetry of the problem.

\subsection{Stationary states: Two-layers polarization}
\begin{figure}
    \centering
    \begin{subfigure}[b]{0.45\textwidth}
     \centering
    \includegraphics[width=\linewidth]{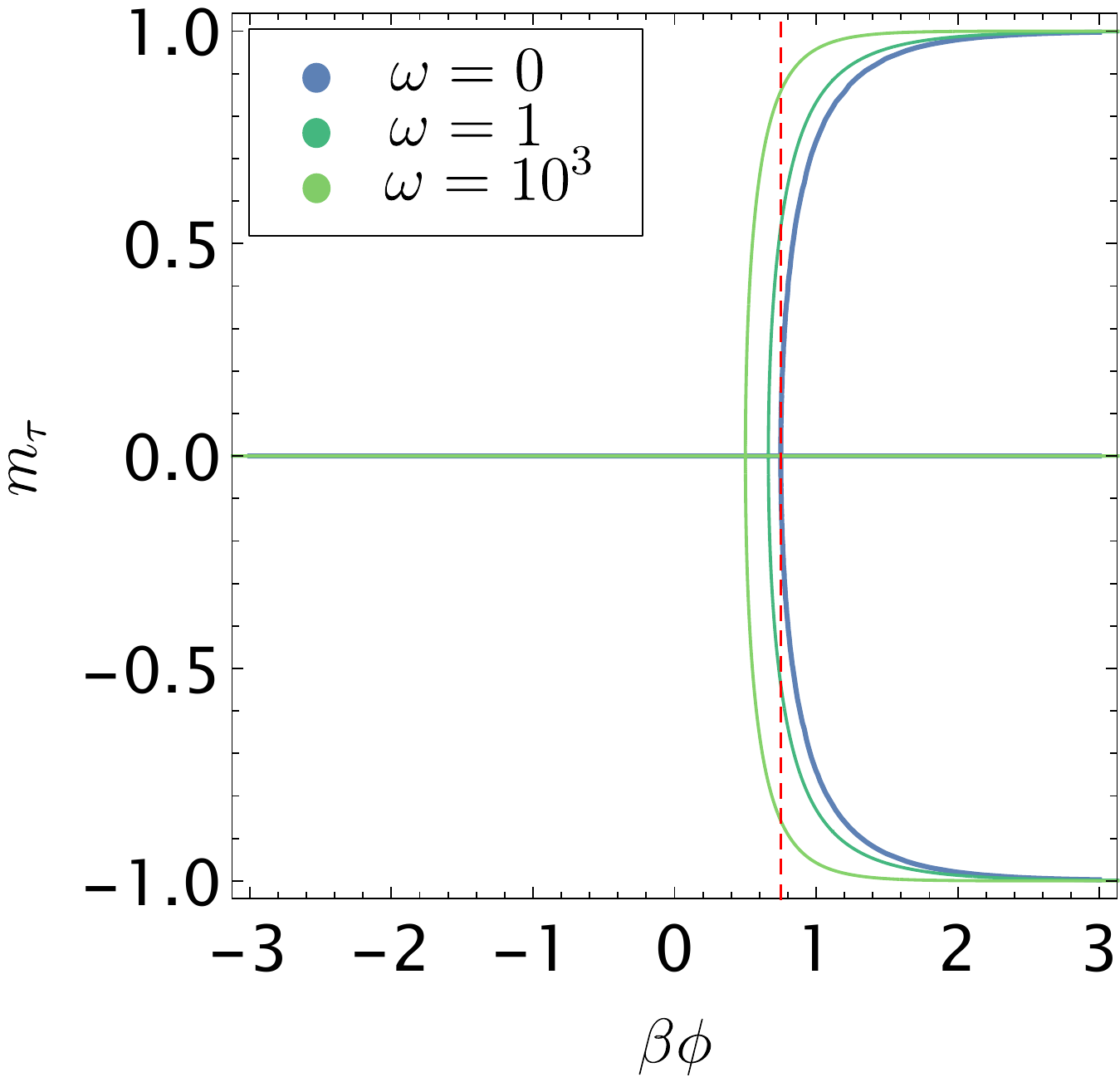}
        \caption{}
    \label{fig:m_general_A}
    \end{subfigure}
    \hfill
    \begin{subfigure}[b]{0.45\textwidth}
     \centering
    \includegraphics[width=\linewidth]{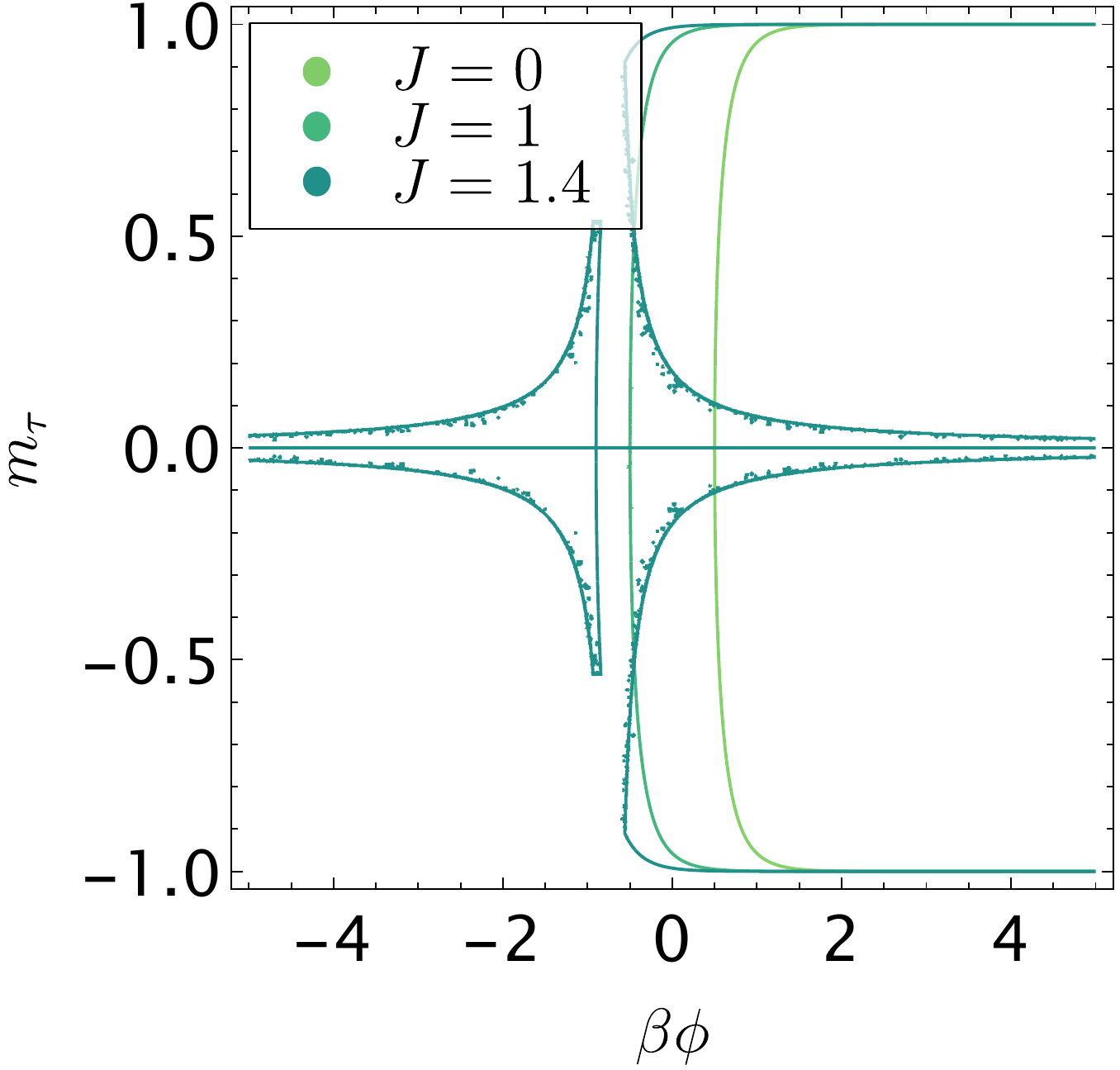}
        \caption{}
    \label{fig:m_general_B}
    \end{subfigure}
\caption{Solution for the magnetization $m_\tau$ as function of $\beta\phi$, when the couplings are set to (a) $J=0$ for various $\omega$ and (b) $\omega=10$ for various $J$. The red dashed line in (a) outlines the transition point in the limiting case $\omega=J=0$.}
\label{fig:m_general}
\end{figure}

When all the couplings are non-zero, the system displays a rich behaviour. 
As our earlier analysis revealed that when $m_\tau=0$ then $m_\sigma=0$, we show, in figure \ref{fig:m_general}, the bifurcation diagram, from the trivial solution, of $m_\tau$, when varying $\omega$ at fixed $J$ (panel a) and varying $J$ at fixed $\omega$ (panel b). The latter shows the emergence of different branches that are expected to be stable at different temperatures. 
The non-zero solution of the system \eqref{eq:magn_state} can be obtained analytically in the low temperature limit, where disorder is low and consensus emerges. This limit is found by extracting the leading order in both numerator and denominator of equations \eqref{eq:magn_state} when $\beta\rightarrow \infty$. The findings of this expansion are summarised in Table \ref{tab:magn_zeroT}.

\begin{table}[ht]
\caption{\label{tab:magn_zeroT}Table showing the non-zero solutions for the magnetization in the $\beta \rightarrow \infty$ regime.}
\begin{tabular}{ c || c }
\hline 
$(m_\sigma^\star,m_\tau^\star)$ &  domain \\ \hline      
 & $\omega + J +2\phi >0$ \\
$\pm(1,1)$ & $J +\omega >0$\\
& $J+\phi>0$ \\
\hline
 & $\omega + J -2\phi >0$  \\
$\pm(1,-1)$ & $J +\omega <0$ \\
 & $J-\phi>0$ \\
\hline  
\end{tabular}
\end{table}
The two domains for the parallel and anti-parallel solutions are disjoint, suggesting that the choice of the parameters determines the agreement of polls and election result, (at least) for the case of $\beta\rightarrow \infty$.

The result of the low-temperature analysis is twofold: firstly, it shows the existence of a region for the triplet of parameters $(J,\omega,\phi)$ in which agreement between declared and true vote is reached. This being, not surprisingly, the region where all the parameters are positive, corresponding to the scenario in which there is a tendency to both convert and conform and be truthful to one's own opinion in the ballot box. Moreover, agreement can occur when there seemingly is a tendency to appear in dissent with what others say $(\phi<0)$ but alongside a drive to convert one's real belief to what is publicly expressed $(J>0,\ \omega>0)$, and with a low tendency to lie or deceive. Secondly, the result shows the emergence of disagreement for some values of $(J,\omega,\phi)$, corresponding to election results overturning the poll predictions (which has occurred for  Bradley's election results). In a positive conversion strength setting $(J>0)$, this may occur when there is a strong tendency to conceal $(\omega<0)$. The case $J<0$ is the mirroring scenario to what we have just described and disagreement emerges for a wider range of $\omega$ and $\phi$.

\begin{figure}
    \centering
    \begin{subfigure}[b]{0.45\textwidth}
     \centering
    \includegraphics[width=\textwidth]{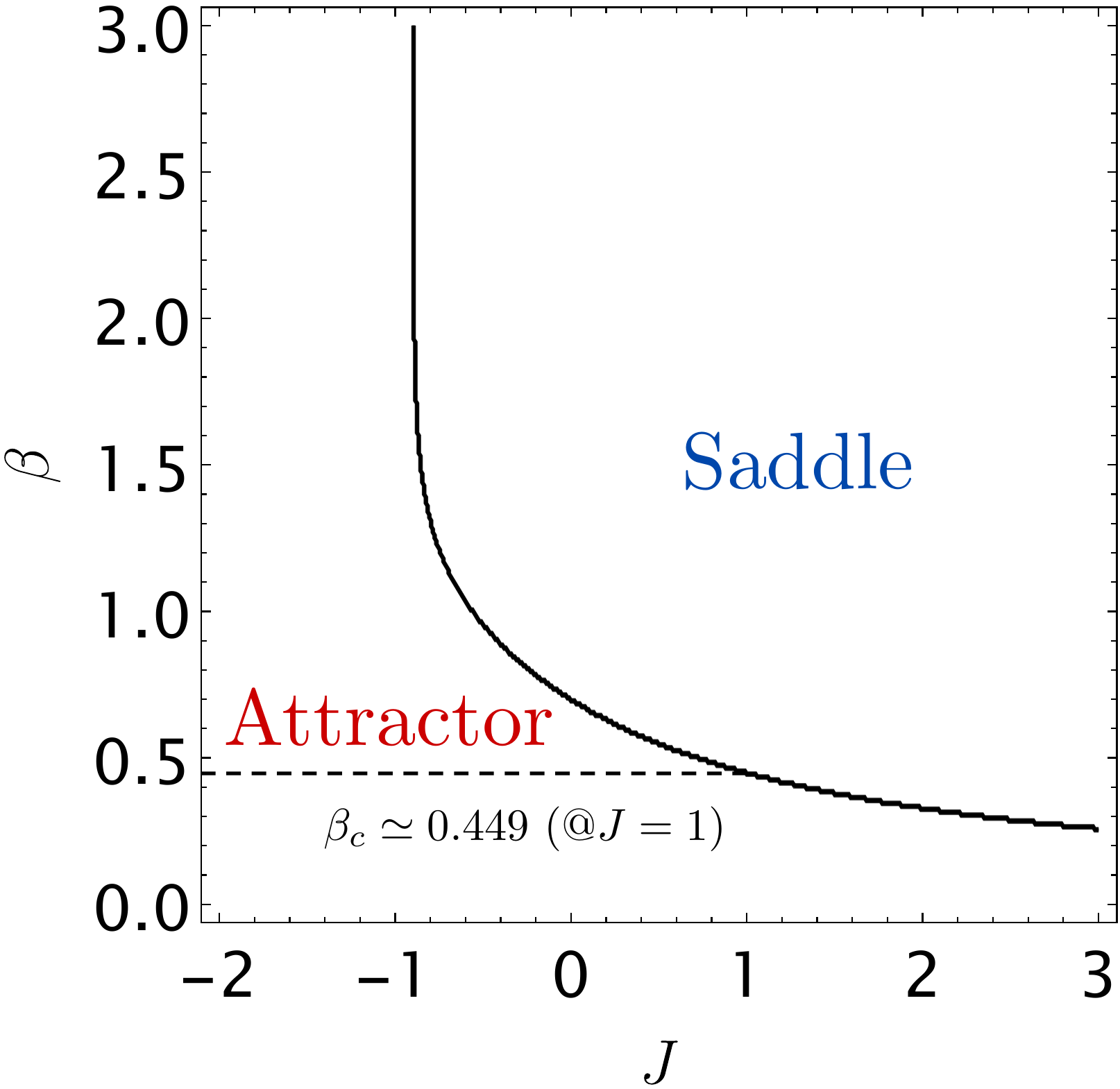}
    \caption{}
    \label{fig:betaJ00sol_A}
    \end{subfigure}
    \hfill
    \begin{subfigure}[b]{0.45\textwidth}
     \centering
    \includegraphics[width=\textwidth]{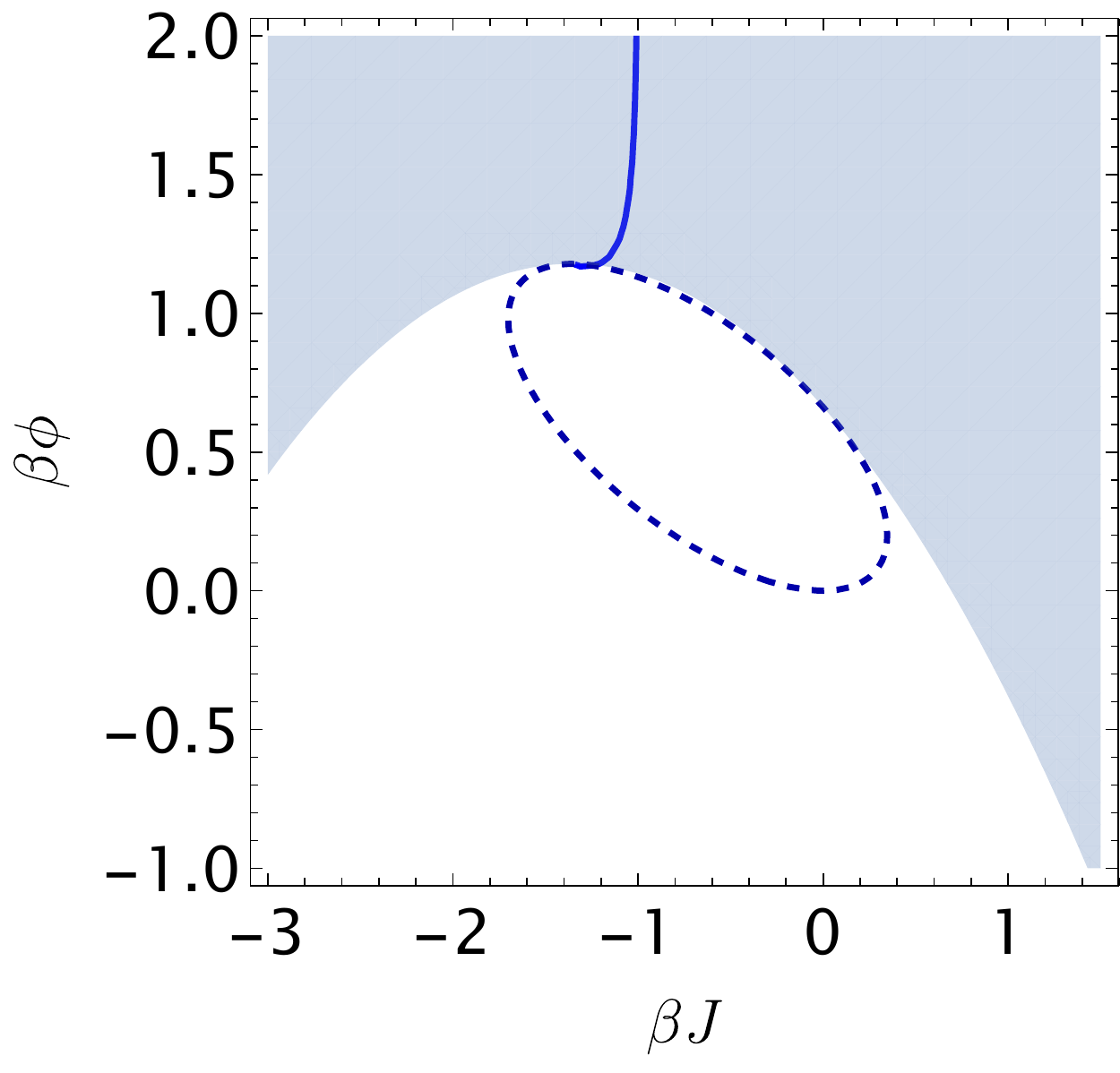}
    \caption{}
    \label{fig:betaJ00sol_B}
    \end{subfigure}    
	\caption{Stability regions of the solution $(m_\sigma,m_\tau)$ as function of $\beta$ and $J$, for fixed parameters $\omega=\phi=1$ (a) and, in the parameter space $\beta J-\beta \phi$ at $\beta\omega=1$. In particular, (a) shows the  transition line between the stability regions of the paramagnetic solution $(0,0)$ . The shaded area in (b) corresponds to a stable polarization in both layers. The blue line corresponds to polarization only on the visible layer.}
	\label{fig:betaJ00sol}
\end{figure}
At finite temperature, one can check the stability of the numerical solution of the system of equations \eqref{eq:magn_state} $m_\sigma^*$ and $m_\tau^*$ by linearizing the system, as described in the previous paragraph. Keeping the couplings fixed, one may investigate when the transition occurs as a function of temperature. In figure \ref{fig:betaJ00sol_A} couplings are fixed to $\omega=\phi=J=1$, the critical temperature is found to be $\beta_c\simeq 0.449$. Figure \ref{fig:betaJ00sol_B} shows the stability of the non trivial solutions in the effective couplings space (with $\beta\omega = 1$), with the light blue region corresponding to stable solutions and, in particular, the blue line corresponding to the stable one-layer polarization solution, as seen in the previous paragraph.

\subsection{Voters self-alignment and correlation - The effect of the temperature}\label{sec:self_align}
An interesting observable is the voting self-alignment, defined as 
\begin{equation}
    s = \frac{1}{N}\sum_i \sigma_i \tau_i\ ,
\end{equation}
which takes values in $[-1,1]$ and describes a voter's tendency to be truthful ($s=1$) or lie ($s=-1$) about their hidden preference, with $s=0$ indicating an erratic behaviour where the voter's preference is uncorrelated with the declared one. We will see that the temperature has a peculiar effect on this local spin correlator.

\begin{figure}[ht]
    \centering
    \includegraphics[width=0.5\textwidth]{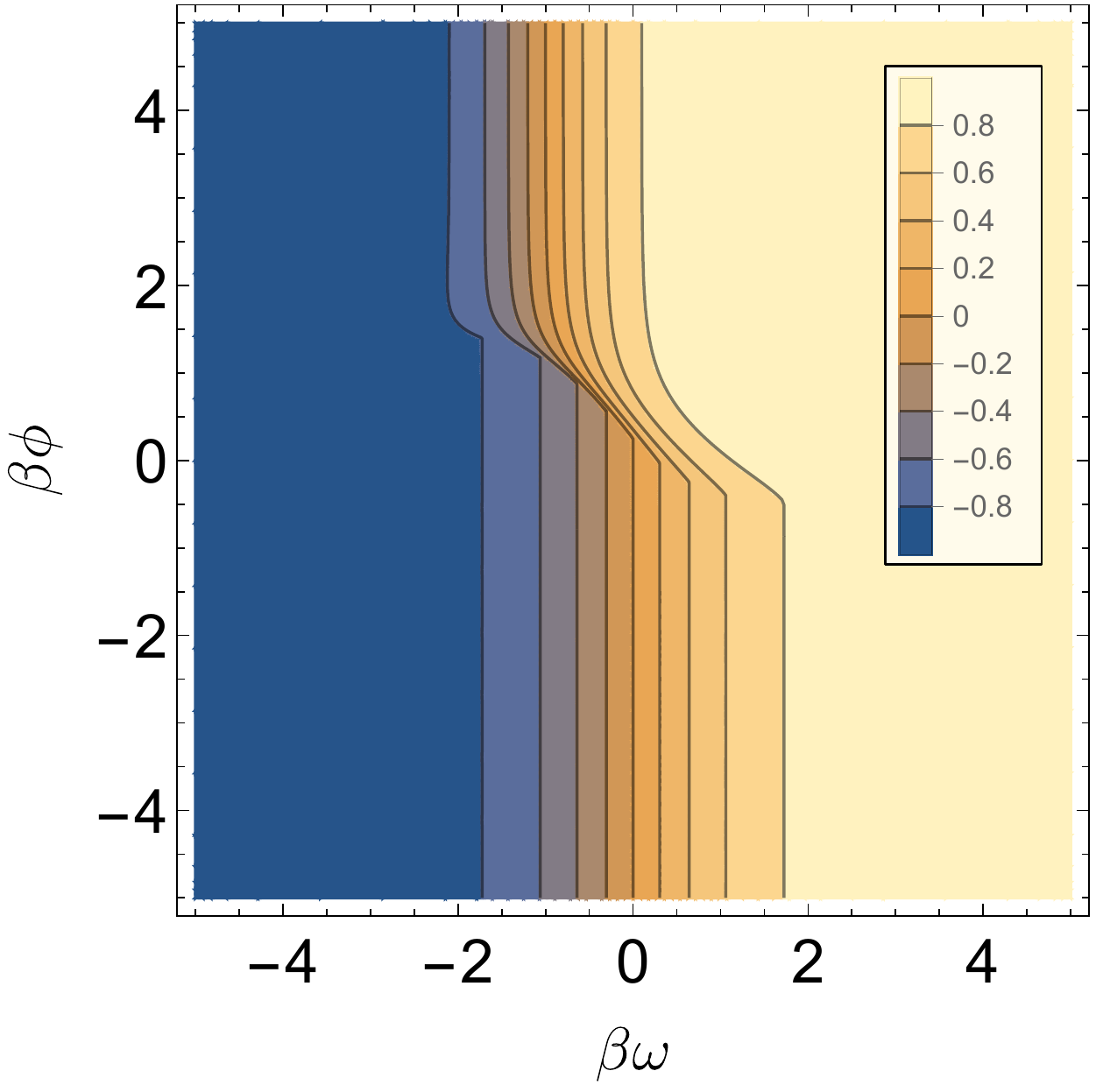}
\caption{Self-alignment $\langle s \rangle$ as a function of the parameters $\beta\omega$ and $\beta\phi$, with fixed $J=1$.}
\label{fig:align}
\end{figure}

\begin{figure}
    \centering
    \begin{subfigure}[b]{0.45\textwidth}
     \centering
    \includegraphics[width=\textwidth]{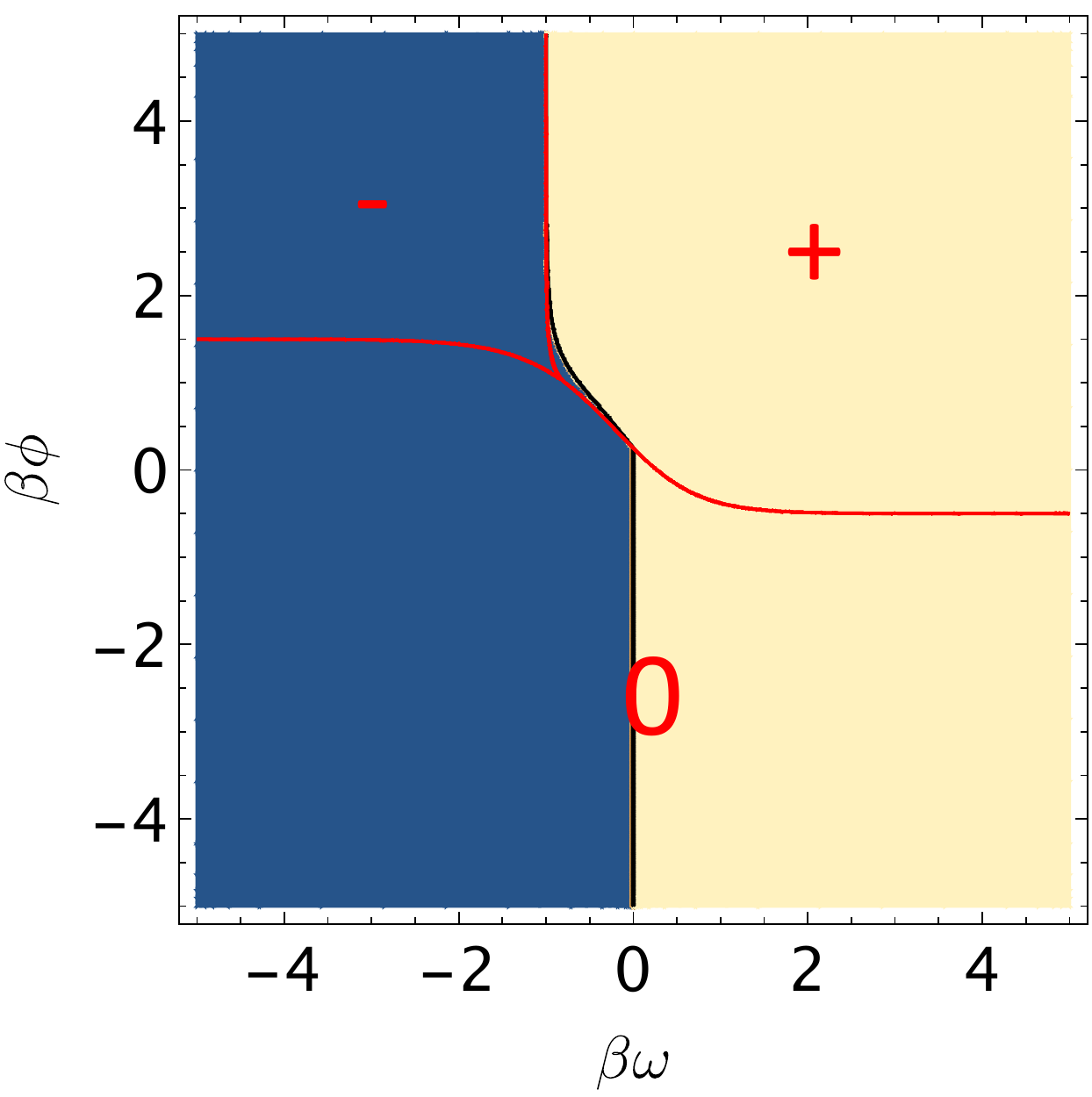}
        \caption{}
    \label{fig:sign_A}
    \end{subfigure}
    \hfill
    \begin{subfigure}[b]{0.45\textwidth}
     \centering
    \includegraphics[width=\textwidth]{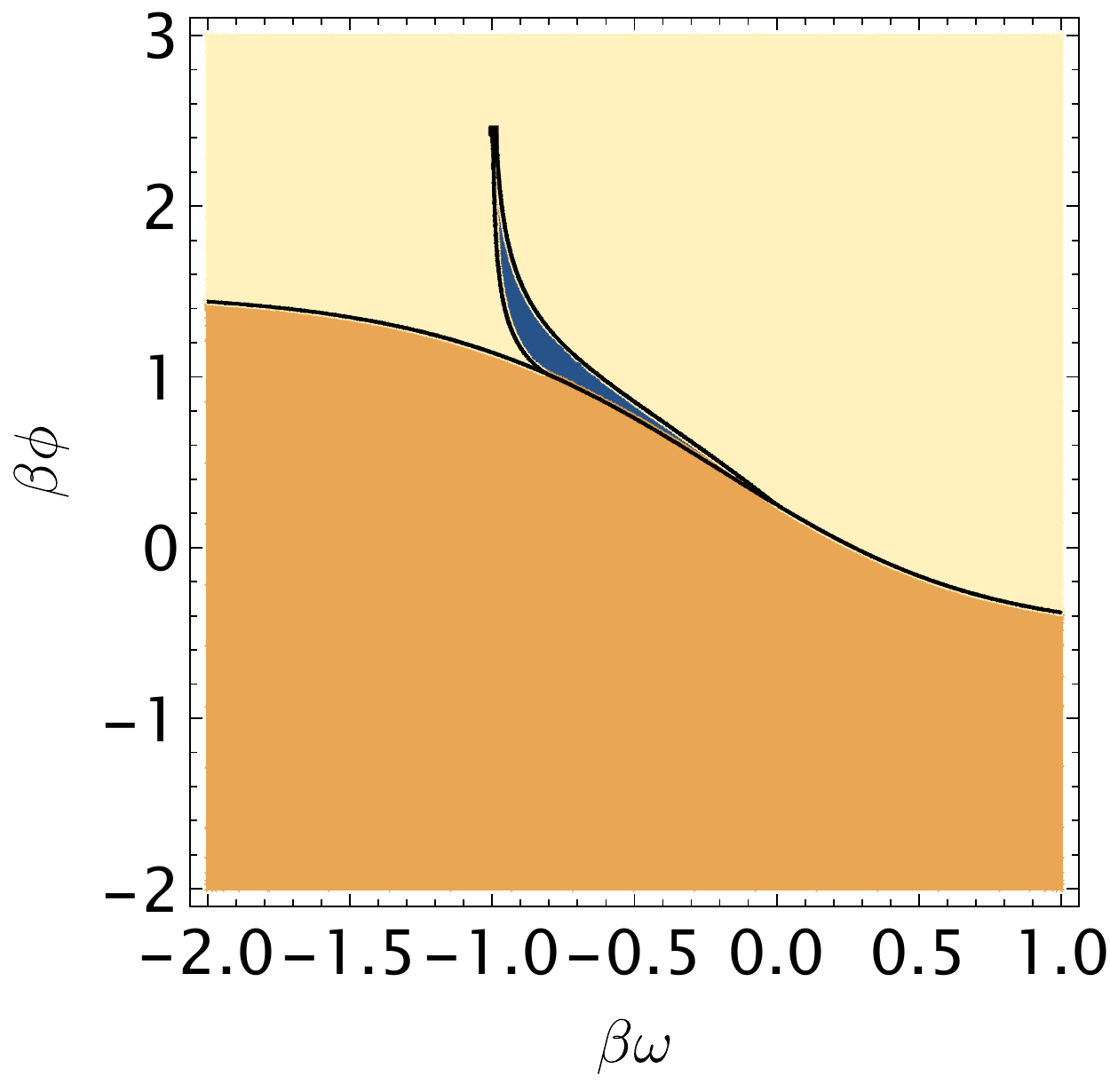}
        \caption{}
    \label{fig:sign_B}
    \end{subfigure}    
\caption{Comparison between the sign of the product of the saddle-point magnetizations $m_\tau^\star m_\sigma^\star$ and the saddle-point self-alignment $\langle s \rangle$ as a function of the parameters $\omega$ and $\phi$, with fixed $J=\beta=1$. 
The red curve in (a) separates the regions where $m_\tau^\star m_\sigma^\star$ is positive, negative or zero whereas the blue and yellow areas indicate the regions where the self-alignment $\langle s\rangle$ is negative and positive, respectively.
Plot (b) is derived from (a) by multiplying the signs of both quantities and zooming around the ``disagreement'' area, highlighted in blue.}
\label{fig:sign}
\end{figure}

The expression for the average  $\langle s \rangle$ can be directly found by differentiating the free energy density in equation \eqref{eq:free_energy} with respect to the conjugated variable $\omega$
\begin{align}
     \langle s \rangle &= \frac{1}{\beta N}\frac{\partial}{\partial\omega}\log Z_N\nonumber\\
    &= 2 \frac{e^{\beta J m_\tau^\star}\sinh\left(\beta \omega+\beta J m_\sigma^\star +2\beta \phi m_\tau^\star\right)-e^{-\beta J m_\tau^\star}\sinh\left(-\beta \omega+\beta J m_\sigma^\star +2\beta \phi m_\tau^\star\right)}{e^{\beta J m_\tau^\star}\left(1+2\cosh\left(\beta \omega+\beta J m_\sigma^\star +2\beta \phi m_\tau^\star\right)\right)+e^{-\beta J m_\tau^\star}\left(1+2\cosh\left(-\beta \omega+\beta J m_\sigma^\star +2\beta \phi m_\tau^\star\right)\right)}\ .
\label{eq:avg_s}
\end{align}
A plot of the average alignment as a function of the model parameters can be seen in figure \ref{fig:align}. Here one can appreciate the effect of $\omega$ on the alignment of the system: as expected, the tendency of voters to lie (or be truthful)  ultimately controls the correlation between the two spin layers.
When comparing the sign of the saddle-point solution for the magnetizations and the self-alignment, figure \ref{fig:sign}, we observe the existence of a ``disagreement area'' in which, on average, each voter tends to declare the opposite of their true intention, however, the two layers tend to align, i.e. the overall election outcome and poll trend tend to agree on average.
This counter-intuitive effect is induced by the temperature $1/\beta$ which gradually decouples the spins across the two layers and within each layer. 

\begin{figure}[ht]
    \centering
    \includegraphics[width=0.5\textwidth]{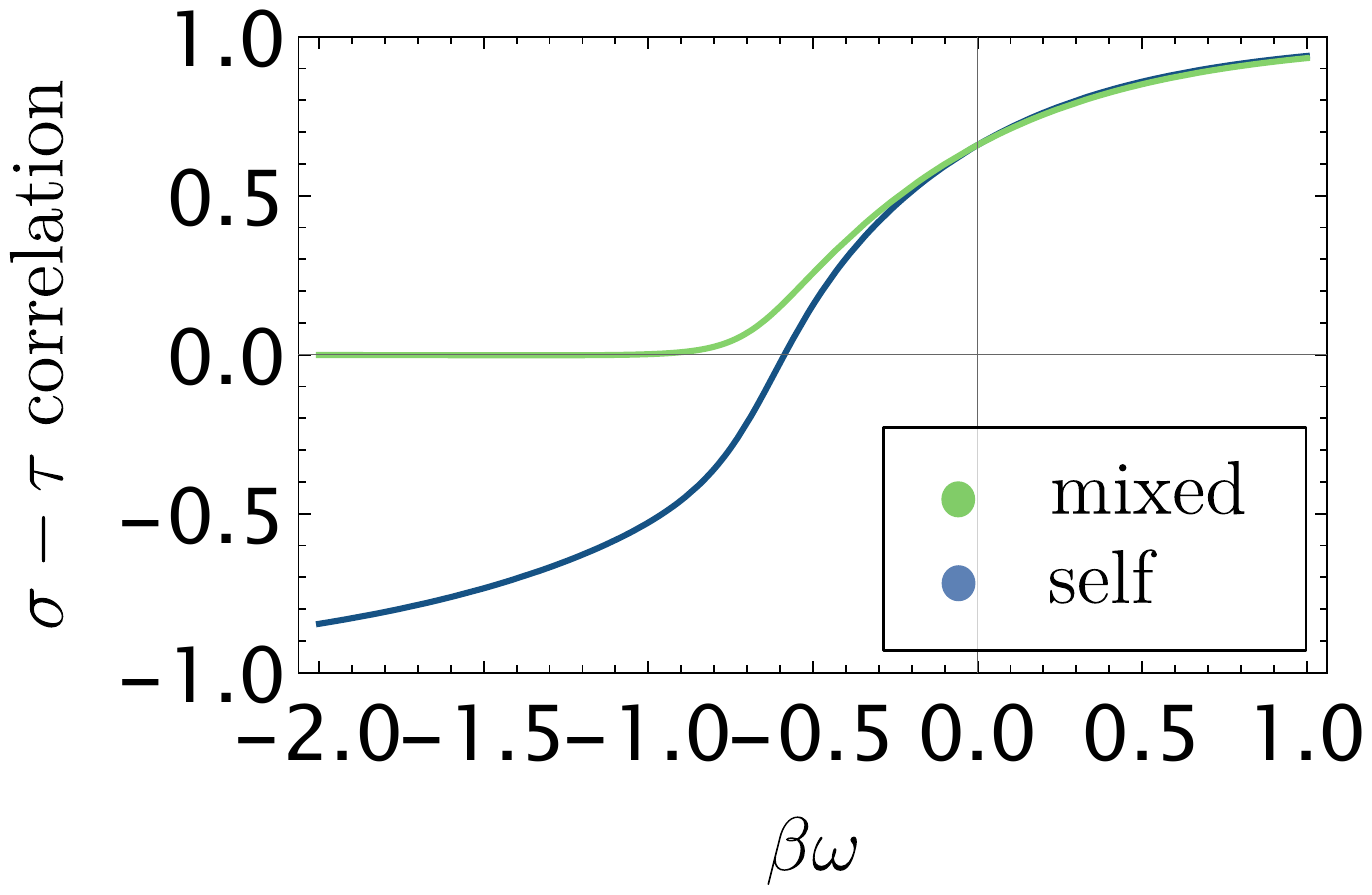}
\caption{Mixed and self voters correlator, blue and green curve respectively, as a function of $\beta \omega$, for a system with size $N=100$. Parameters are set to $\beta J=\beta \phi=1$. The two correlators have opposite sign in correspondence with the emergence of the disagreement area.}
\label{fig:c_sigmatau}
\end{figure}

We can better appreciate this when analysing the self and mixed voters correlator, namely $\rho_{\sigma\tau}=\frac{1}{N}\sum_i\left(\left<\sigma_i\tau_i \right> - \left<\sigma_i \right> \left<\tau_i \right>\right)$ and $\tilde{\rho}_{\sigma\tau}=\frac{1}{N(N-1)}\sum_{i\neq j}\left(\left<\sigma_i\tau_j \right> - \left<\sigma_i \right> \left<\tau_j \right>\right)$. As shown in figure \ref{fig:c_sigmatau}, in the disagreement area, the single voter (self) correlator $\rho_{\sigma\tau}$ is negative while the mixed correlator $\tilde{\rho}_{\sigma\tau}$ becomes positive. The gap between these two curves closes as the temperature decreases and thus this phenomenon gradually disappears. We calculate the expression for the correlators, at finite $N$, in Appendix \ref{sec:correlator}.

\section{Beyond averages}\label{sec:beyond_avg}
The model is conceived to reflect wide availability of surveys information, i.e. the information collected by pollsters being shared with the entire network. This mean-field set up allows us the rare luxury to be able to compute exactly the full distribution of the main observables of the system -- even at finite $N$ -- in terms of integrals that are easily evaluated numerically. In this section we calculate the marginal distribution for the magnetizations in both layers, which are then employed, in the next section, to showcase a possible application of the model. 

Starting with equation \eqref{eq:joint_p}, we perform the Hubbard-Stratonovich transformation to replace the quadratic interaction in $\bm\tau$ with a Gaussian field $\mu$, 
\begin{equation}
P(\bm\sigma,\bm\tau) 
= \frac{1}{Z_N}\int \frac{d\mu}{\sqrt{4\pi/\beta\phi N}}\exp\left(-\frac{\beta\phi N}{4}\mu^2\right)\exp\left\{\frac{\beta J}{N}\sum_{j}\sigma_j\sum_i\tau_i+\beta\omega\sum_{i} \sigma_i\tau_i+\beta\phi\mu\sum_i  \tau_i\right\}\ .
\label{eq:HS_jointp_m}
\end{equation}
The partition function follows by summing over the configurations
\begin{align}
{Z_N}& = \sum_{\{\bm\sigma\}} \int \frac{d\mu}{\sqrt{4\pi/\beta\phi N}}\exp\left(-\frac{\beta\phi N}{4}\mu^2\right)\prod_i\sum_{\{\tau_i\}}\exp\left\{\frac{\beta J}{N}\sum_{j}\sigma_j\tau_i+\beta\omega \sigma_i\tau_i+\beta\phi\mu\tau_i\right\}\nonumber\\
&= \sum_{\{\bm\sigma\}} \int \frac{d\mu}{\sqrt{4\pi/\beta\phi N}}\exp\left(-\frac{\beta\phi N}{4}\mu^2\right)\prod_i\left[1 + 2\cosh \left(\frac{\beta J}{N}\sum_{j}\sigma_j+\beta\omega \sigma_i+\beta\phi\mu \right)\right]\ .
\end{align}
Now, by using the identity $\int \frac{dm_\sigma d\xi}{2\pi/N} e^{-\I N\xi(m_\sigma -\frac{1}{N}\sum_{i=1}^N \sigma_i)} = \int dm_\sigma \delta\left(m_\sigma-\frac{1}{N}\sum_{i=1}^N \sigma_i\right)=1$, allows replacing $\frac{1}{N}\sum_{i=1}^N \sigma_i$ with $m_\sigma$ so we are able to perform the expectation and calculate a more explicit form of $Z_N$
\begin{align}
Z_N &= \int \frac{d\mu}{\sqrt{4\pi/\beta\phi N}}e^{-\frac{\beta\phi N}{4}\mu^2}\int \frac{dm_\sigma d\xi}{2\pi/N}\prod_i \sum_{\sigma_i}\left[\left(1 + 2\cosh F_{\sigma_i}(m_\sigma)\right)e^{\I\xi\sigma_i}\right]e^{-\I m_\sigma \xi N}\nonumber\\
&= \sum_{n=0}^N\binom{N}{n}\int \frac{d\mu}{\sqrt{4\pi/\beta\phi N}}e^{-\frac{\beta\phi N}{4}\mu^2} \left(1 + 2\cosh F_{+1}\left(\frac{N-2n}{N}\right)\right)^{N-n} \left(1 + 2\cosh F_{-1}\left(\frac{N-2n}{N}\right)\right)^n \ ,
\end{align}
with $F_{\sigma_i}(m_\sigma)=\beta J m_\sigma+\beta\omega\sigma_i +\beta\phi\mu$.
Introducing the Fourier representation of Dirac delta, the distribution for the magnetization in the 
visible layer is found in a similar way to the partition function 
\begin{align}
    P(m_\tau) &= \sum_{\bm\sigma,\bm\tau}P(\bm\sigma,\bm\tau) \delta\left(m_\tau - \frac{1}{N}\sum_i \tau_i\right)\nonumber\\
    &= \frac{1}{Z_N}\int \frac{d\mu}{\sqrt{4\pi/\beta\phi N}}e^{-\frac{\beta\phi N}{4}\mu^2}\int\frac{dm_\sigma d\xi d\zeta}{(2\pi/N)^2} \sum_{n=0}^{N}\binom{N}{n} \left[\left(e^{ F_{+1}(m_\sigma)} e^{\I\xi}+e^{F_{-1}(m_\sigma)} e^{-\I\xi}\right)e^{\I\zeta} + \left( e^{\I\xi}+e^{-\I\xi}\right)\right]^{N-n} \times  \nonumber\\
    &\quad \times \left(e^{-F_{+1}(m_\sigma)} e^{\I\xi}+e^{-F_{-1}(m_\sigma)} e^{-\I\xi}\right)^n e^{-\I\zeta n} e^{-\I N(\xi m_\sigma +\zeta m_\tau)}\ .
\label{eq:P_mtau}
\end{align}
Although the integrals above can be computed explicitly by simply applying the binomial expansion, we prefer to leave their integral expressions here as they are more stable to evaluate numerically.

\subsection{Maximum likelihood estimation of the true vote}\label{sec:ML_ms}
In this final section we discuss an application of our model to election prediction. 
When a prior belief $P(\bOmega)$ can be constructed for the set of couplings $\bOmega=(\phi, \omega, J)$, one may compute the likelihood of $m_\sigma$ given the observed $m_\tau$
\begin{equation}
    P(m_\sigma|m_\tau)=\int d\bOmega P(\bOmega) \frac{P(m_\sigma,m_\tau|\bOmega)}{P(m_\tau|\bOmega)} \ .
    \label{eq:cond_inference}
\end{equation}
If $\bOmega$ is known exactly, i.e. $P(\bOmega)=\delta(\bOmega-\bOmega^*)$, then 
the
conditional probability of the true votes magnetization, given the average vote declaration, is simply given by
$P(m_\sigma|m_\tau)=P(m_\sigma,m_\tau)/P(m_\tau)$.
In order to get the maximum-likelihood (ML) estimation for the concealed votes $\hat{m}_\sigma$ as a function of the observed $m_\tau$, we maximize the log-likelihood (i.e. the logarithm of the conditional distribution) with respect to $m_\sigma$. This is equivalent to maximizing the 
logarithm of the joint distribution, 
since the marginal $P(m_\tau)$ in equation \eqref{eq:P_mtau} is independent of the concealed average vote. The joint distribution is found to be (see appendix \ref{sec:P_m_sigma_m_tau} for more details)
\begin{align}
    P(m_\sigma,m_\tau) 
&= \frac{1}{Z_N}\int \frac{d\mu}{\sqrt{4\pi/\beta\phi N}}e^{-\frac{\beta\phi N}{4}\mu^2}\int\frac{d\xi d\zeta}{(2\pi/N)^2}  \left[\left(e^{ F_{+1}(m_\sigma)} e^{\I\xi}+e^{F_{-1}(m_\sigma)} e^{-\I\xi}\right)e^{\I\zeta} +
2\cos \xi+ \right. \nonumber\\
&\quad  \left. +\left(e^{-F_{+1}(m_\sigma)} e^{\I\xi}+e^{-F_{-1}(m_\sigma)} e^{-\I\xi}\right)e^{-\I\zeta}\right]^N e^{-\I N(\xi m_\sigma +\zeta m_\tau)}\ .
\end{align}

\begin{figure}[ht]
    \centering
    \includegraphics[width=0.5\textwidth]{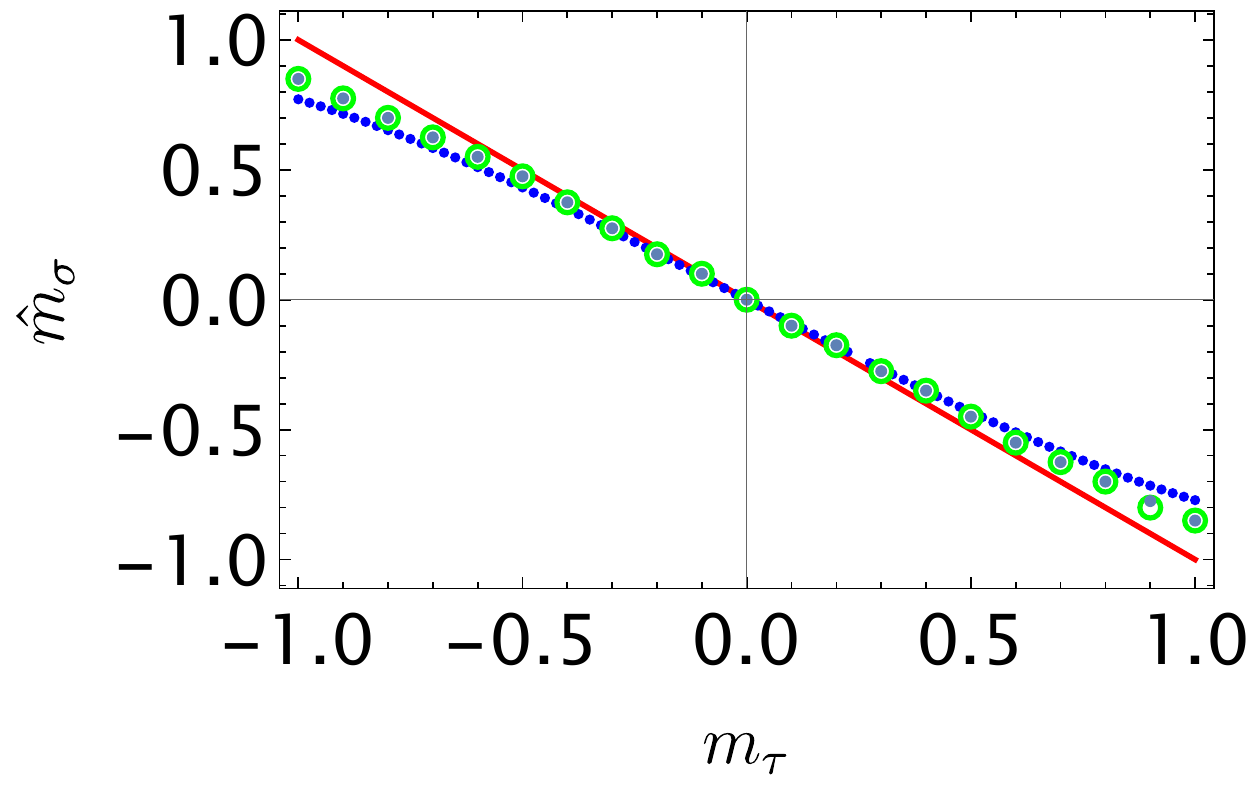}
\caption{Maximum-likelihood (ML) estimate $\hat{m}_\sigma$ of the true vote as a function of the average declared vote $m_\tau$  in a system with size $N=100$. The red line represents $\hat{m}_\sigma = -m_\tau$. 
The blue dotted line shows $\hat{m}_\sigma$ in the case of a delta prior distribution $P(\bOmega)=\delta(\bOmega-\bOmega^*)$, peaked on the parameter values
$\bOmega^* = (J^*,\omega^*,\phi^*)=(-1,-1,1)$ and $\beta=0.5$. The grey and green dots show $\hat{m}_\sigma$ in the case of a Gaussian prior $P(\bOmega)=\mathcal{N}(\bOmega^*,\Sigma_\Omega)$ with covariance matrix $\Sigma_\Omega=\alpha\mathbb{1}$, with $\mathbb{1}$ the $3\times 3$ identity matrix, respectively with $\alpha=0.8$ and $\alpha=1.6$.}
\label{fig:ML_ms}
\end{figure}

Figure \ref{fig:ML_ms} (blue dots) shows a numerical solution of $\hat{m}_\sigma$, for a fixed value of the couplings $J=\omega=-1$, $\phi=1$ and $\beta=0.5$. These values for the couplings realize an outcome $\hat{m}_\sigma \sim -m_\tau$, i.e., the true vote outcome is expected to be opposite to the declared one: it is a Bradley effect-like scenario in which the voters, by concealing their true beliefs, would crown a candidate that was behind the other one in polls.
One can also get the distribution $P(m_\sigma|m_\tau)$ when there is a small uncertainty around $\bOmega^*$, i.e. $P(\bOmega)\neq \delta(\bOmega-\bOmega^*)$. 
We demonstrate this for a Gaussian distribution 
$P(\bOmega)$, with diagonal covariance matrix $\Sigma_\Omega=\text{diag}(\sigma_{\phi},\sigma_{\omega},\sigma_{J})$, by solving the integral in equation \eqref{eq:cond_inference}. We compare this case with a delta-prior in figure \ref{fig:ML_ms} in which the gray dots correspond to a Gaussian prior with  $\sigma_{\phi}=\sigma_{\omega}=\sigma_{J} = 0.8$ and, similarly, the green circles to $\sigma_{\phi}=\sigma_{\omega}=\sigma_{J} = 1.5$.

Finally, the predictions of this model can be improved by refining $P(\bOmega)$ with knowledge of historical data, i.e. defining the dataset 
$D=\{(m_\tau^1, m_\sigma^1),\dots,(m_\tau^P, m_\sigma^P)\}$ of past associations between declarations and election outcomes, one can compute 
\begin{equation}
P(\bOmega|D)=\frac{P(\bOmega) \prod_{\mu=1}^P P(m_\sigma^\mu|m_\tau^\mu,\bOmega)}{\int d\bOmega P(\bOmega) \prod_{\mu=1}^P P(m_\sigma^\mu|m_\tau^\mu,\bOmega)}\ ,
\label{eq:distr_couplings}
\end{equation}
and predict the future outcome of an election $m_\sigma$ with declaration $m_\tau$ (not part of the historical dataset) as 
\begin{equation}
    P(m_\sigma|m_\tau,D)=\int d\bOmega P(\bOmega|D)P(m_\sigma|m_\tau,\bOmega)\ .
    \label{eq:distr_ms}
\end{equation}
Applying equations \eqref{eq:distr_couplings} and \eqref{eq:distr_couplings} to a real election data-set goes beyond the scope of this work, which is instead focused on the static analysis of the interacting two-layers system of spins. We intend to exploit the machinery introduced in this section, in a forthcoming data-oriented work.

\section{Conclusion and Outlook}\label{sec:conclusion}
Understanding biases in decision-making is a central topic in the present era in which information can be easily harvested but equally easily manipulated. Biases get particularly evident when it comes to election results and they have been attracting much interest from both pollsters and sociologists. Elections are in fact a unique opportunity for many people to express their true preferences and beliefs, which then also become widely available to the general public in aggregate (anonymous) form. Understanding biases in this setting grants the sociologist and the political scientists insights into the population stance on the most pressing societal issues. 

In our work we proposed a two-layer system of interacting spins, representing the concealed and expressed opinions of a population of $N$ individuals. We explicitly model the interaction between these two layers in the form of the interaction Hamiltonian \eqref{eq:interact_H} with a conversion and a compliance coupling, and we further introduce a social integrity term that takes into account one own’s level of (moral) integrity. We also allow a subset of the visible layer not to be sampled, because either the interviewees refuse to declare their preference or they are not polled at all.

We study the system in equilibrium in the setting in which declared votes are known to the entire population (fully connected graph in the visible layer) and there is a symmetric interaction between one's own opinion and these public declarations. The hidden layer is internally disconnected, as we assume that the true opinion be concealed. We show that this setting allows the system to polarize, in both layers, for some range of values for the couplings and social temperature. 

In section \ref{sec:model}, we study the bipartite system in the thermodynamic limit and derive the set of coupled equations for the magnetization in equation \ref{eq:magn_state}. In section \ref{sec:eq_analisys} we investigate the region of values for the couplings for which the polarization arises, particularly at high and low social temperature, and study the stability of the polarization, as presented in figure \ref{fig:betaJ00sol}.
We then discuss the effect of the social temperature on the correlation (or vote alignment) which causes the emergence of a “self-disagreement” zone (in the couplings space), for which one's own true and declared votes are on average anti-correlated, but the resulting true vote agrees with the public declarations, as highlighted in figure \ref{fig:sign}. This mismatch can be explained by the tendency of voters to conform to (or
be converted by) their perception of their peers’ declared votes which are, however, made fuzzy by the temperature.
Finally, in Section \ref{sec:ML_ms}, we briefly present an application of the model on election prediction, assuming prior knowledge of 
the coupling strengths for a given population.

An interesting pathway for future work would be the inference of the couplings from real data. 
Such modelling approach would complement an emerging strand of literature that focuses on machine learning methods aimed at quantifying biases in political decision making and produce better election predictions \cite{ZSCCM21}.

Besides inference, it would be interesting to analyse the dynamics of our model and study the convergence time to a steady state. This would allow us to relax the assumption about the symmetry of the interactions between the two sets of spins. In our equilibrium setting, to ensure that the detailed balance condition is satisfied, the hidden variable set $\bm \sigma$ influences the full $\bm \tau$ so that true opinions can be seen by everyone and thus influence the expressed opinions. Dropping this constraint would produce a more realistic dynamical voting model. 

Finally, while our model is focused on election settings, we note that it could equally well apply to any circumstances in which a public and concealed variable become available, or can be somehow inferred, for instance, in the context of improving self-reporting in sports, or correcting biases in clinical trials. 

%%%%%%%%%%%%%%%%%%%%%%%%%%%%%%%%%%%%%%%%%%%%%
%%%%%%%%%%%%%%%%%%%%%%%%%%%%%%%%%%%%%%%%%%%%%
\begin{acknowledgments}
%%%%%%%%%%%%%%%%%%%%%%%%%%%%%%%%%%%%%%%%%%%%%
%%%%%%%%%%%%%%%%%%%%%%%%%%%%%%%%%%%%%%%%%%%%%
PV and ET acknowledge support from UKRI Future Leaders Fellowship scheme [n. MR/S03174X/1]. Y-PF is supported by the EPSRC Centre for Doctoral Training in Cross-disciplinary Approaches to Non-Equilibrium Systems (CANES EP/L015854/1).
\end{acknowledgments}

%\bibliography{BE_bib}

%%%%%%%%%%%%%%%%%%%%%%%%%%%%%%%%%%%%%%%
%      BIBLIOGRAPHY
%%%%%%%%%%%%%%%%%%%%%%%%%%%%%%%%%%%%%%%
%

\section*{Appendix}
\appendix

\subsection{\texorpdfstring{$\sigma - \tau$}{}  correlation}\label{sec:correlator}
The correlation between two spins of different species $\sigma_m$ and $\tau_k$ can be calculated as follows.
Starting with equation \eqref{eq:joint_p}, we perform the Hubbard-Stratonovich transformation to replace the quadratic interaction in $\bm\tau$ with a Gaussian field $\mu$, 
\begin{align}
\left<\sigma_m \tau_k\right> 
&= \frac{1}{Z_N}\int \frac{d\mu}{\sqrt{4\pi/\beta\phi N}}e^{-\frac{\beta\phi N}{4}\mu^2} 
\sum_{\{\bm\sigma\}}\sigma_m\sum_{\{\bm\tau\}}\tau_k\prod_{i}\exp\left\{\frac{\beta J}{N}\sum_{j}\sigma_j\tau_i+\beta\omega \sigma_i\tau_i+\beta\phi\mu\tau_i\right\}\nonumber\\
&= \frac{1}{Z_N}\int \frac{d\mu}{\sqrt{4\pi/\beta\phi N}}e^{-\frac{\beta\phi N}{4}\mu^2}\int\frac{dm_\sigma d\hat{m}}{2\pi/N} \sum_{\{\bm\sigma\}}\sigma_m\prod_{i\neq k}\left[\left(1 + 2\cosh F_{\sigma_i}(m_\sigma)\right)e^{\I\hat{m}\sigma_i}\right]\times \nonumber\\
&\quad\times 2\sinh F_{\sigma_k}(m_\sigma)e^{\I\hat{m}\sigma_k} e^{-\I m_\sigma \hat{m}N}\ ,
\end{align}
where $F_{\sigma_i}(m_\sigma)=\beta J m_\sigma+\beta\omega\sigma_i +\beta\phi\mu$. Now,
\begin{itemize}
    \item if $k=m$,
\end{itemize}
\begin{align}
&= \frac{1}{Z_N}\int \frac{d\mu}{\sqrt{4\pi/\beta\phi N}}e^{-\frac{\beta\phi N}{4}\mu^2}\int\frac{dm_\sigma d\hat{m}}{2\pi/N} \left(\left[\left(1 + 2\cosh F_{+}(m_\sigma)\right)e^{\I\hat{m}}\right] + \left[\left(1 + 2\cosh F_{-}(m_\sigma)\right)e^{-\I\hat{m}}\right]\right)^{N-1} \times\nonumber\\
&\quad \times \left(2\sinh F_{+}(m_\sigma)e^{\I\hat{m}} - 2\sinh F_{-}(m_\sigma)e^{-\I\hat{m}}\right)e^{-\I m_\sigma \hat{m}N}\nonumber\\
&= \frac{1}{Z_N}\int \frac{d\mu}{\sqrt{4\pi/\beta\phi N}}e^{-\frac{\beta\phi N}{4}\mu^2}\int\frac{dm_\sigma d\hat{m}}{2\pi/N} \sum_{n=0}^{N-1}\binom{N-1}{n}e^{\I\hat{m}(N-1-2n)} \left(1 + 2\cosh F_{+}(m_\sigma)\right)^{N-1-n}  \times\nonumber\\
&\quad \times \left(1 + 2\cosh F_{-}(m_\sigma)\right)^{n} \left(2\sinh F_{+}(m_\sigma)e^{\I\hat{m}} - 2\sinh F_{-}(m_\sigma)e^{-\I\hat{m}}\right)e^{-\I m_\sigma \hat{m}N}\nonumber\\
&= \frac{1}{Z_N}\sum_{n=0}^{N-1}\binom{N-1}{n} \frac{1}{Z_N}\int \frac{d\mu}{\sqrt{4\pi/\beta\phi N}}e^{-\frac{\beta\phi N}{4}\mu^2}\Bigg[\int dm_\sigma \delta\left(m_\sigma - \frac{N-2n}{N}\right) \left(1 + 2\cosh F_{+}(m_\sigma)\right)^{N-1-n}\times\nonumber\\
& \quad\times \left(1 + 2\cosh F_{-}(m_\sigma)\right)^{n} 2\sinh F_{+}(m_\sigma) -\int dm_\sigma \delta\left(m_\sigma - \frac{N-2-2n}{N}\right) \left(1 + 2\cosh F_{+}(m_\sigma)\right)^{N-1-n}\times \nonumber\\
&\quad \times  \left(1 + 2\cosh F_{-}(m_\sigma)\right)^{n}2\sinh F_{-}(m_\sigma) \Bigg]
\label{eq:corr_self}
\end{align}

\begin{itemize}
    \item if $k\neq m$,
\end{itemize}
\begin{align}
&= \frac{1}{Z_N}\int \frac{d\mu}{\sqrt{4\pi/\beta\phi N}}e^{-\frac{\beta\phi N}{4}\mu^2}\int\frac{dm_\sigma d\hat{m}}{2\pi/N} \left(\left[\left(1 + 2\cosh F_{+}(m_\sigma)\right)e^{\I\hat{m}}\right] + \left[\left(1 + 2\cosh F_{-}(m_\sigma)\right)e^{-\I\hat{m}}\right]\right)^{N-2} \times\nonumber\\
&\quad \times \left(\left(1 + 2\cosh F_{+}(m_\sigma)\right)e^{\I\hat{m}} - \left(1 + 2\cosh F_{-}(m_\sigma)\right)e^{-\I\hat{m}}\right)\left(2\sinh F_{+}(m_\sigma)e^{\I\hat{m}} + 2\sinh F_{-}(m_\sigma)e^{-\I\hat{m}}\right)e^{-\I m_\sigma \hat{m}N}\nonumber\\
&= \frac{1}{Z_N}\int \frac{d\mu}{\sqrt{4\pi/\beta\phi N}}e^{-\frac{\beta\phi N}{4}\mu^2}\int\frac{dm_\sigma d\hat{m}}{2\pi/N} \sum_{n=0}^{N-2}\binom{N-2}{n}e^{\I\hat{m}(N-2-2n)} \left(1 + 2\cosh F_{+}(m_\sigma)\right)^{N-2-n} \times\nonumber\\
&\quad \times \left(1 + 2\cosh F_{-}(m_\sigma)\right)^{n}\left(\left(1 + 2\cosh F_{+}(m_\sigma)\right)e^{\I\hat{m}} - \left(1 + 2\cosh F_{-}(m_\sigma)\right)e^{-\I\hat{m}}\right) \times \nonumber\\
&\quad \times \left(2\sinh F_{+}(m_\sigma)e^{\I\hat{m}} + 2\sinh F_{-}(m_\sigma)e^{-\I\hat{m}}\right)e^{-\I m_\sigma \hat{m}N}\nonumber\\
&= \frac{1}{Z_N}\sum_{n=0}^{N-2}\binom{N-2}{n} \frac{1}{Z_N}\int \frac{d\mu}{\sqrt{4\pi/\beta\phi N}}e^{-\frac{\beta\phi N}{4}\mu^2}\left\{\int dm_\sigma \delta\left(m_\sigma - \frac{N-2n}{N}\right) \left(1 + 2\cosh F_{+}(m_\sigma)\right)^{N-1-n}\times\right.\nonumber\\
& \times \left(1 + 2\cosh F_{-}(m_\sigma)\right)^{n} 2\sinh F_{+}(m_\sigma) +\int dm_\sigma \delta\left(m_\sigma - \frac{N-2-2n}{N}\right) \left[\left(1 + 2\cosh F_{+}(m_\sigma)\right)^{N-1-n} \times \right.\nonumber\\
& \quad \times \left(1 + 2\cosh F_{-}(m_\sigma)\right)^{n}2\sinh F_{-}(m_\sigma) - \left. \left(1 + 2\cosh F_{+}(m_\sigma)\right)^{N-2-n} \left(1 + 2\cosh F_{-}(m_\sigma)\right)^{n+1}2\sinh F_{+}(m_\sigma) \right]+\nonumber\\
& - \left. \int dm_\sigma \delta\left(m_\sigma - \frac{N-4-2n}{N}\right) \left(1 + 2\cosh F_{+}(m_\sigma)\right)^{N-2-n} \left(1 + 2\cosh F_{-}(m_\sigma)\right)^{n+1}2\sinh F_{-}(m_\sigma) \right\}\ .
\label{eq:corr_mxd}
\end{align}
The self and mixed correlators $\rho_{\sigma\tau}=\frac{1}{N}\sum_i\left(\left<\sigma_i\tau_i \right> - \left<\sigma_i \right> \left<\tau_i \right>\right)$ and $\tilde{\rho}_{\sigma\tau}=\frac{1}{N(N-1)}\sum_{i\neq j}\left(\left<\sigma_i\tau_j \right> - \left<\sigma_i \right> \left<\tau_j \right>\right)$ are easily retrieved by summing equations \eqref{eq:corr_self} and \eqref{eq:corr_mxd} over the indices $k$ and $m$, dividing by the appropriate coefficient.

\subsection{Joint distribution for magnetizations}\label{sec:P_m_sigma_m_tau}
Introducing the Fourier representation of Dirac delta, the joint distribution for the
magnetization in the visible and hidden layer can be computed from equation \eqref{eq:joint_p} in the following way
\begin{align}
     P(m_\sigma,m_\tau) 
&= \sum_{\{\bm\sigma,\bm\tau\}}P(\bm\sigma,\bm\tau) \delta\left(m_\tau - \frac{1}{N}\sum_i \tau_i\right)\delta\left(m_\sigma - \frac{1}{N}\sum_i \sigma_i\right)\nonumber\\
&= \frac{1}{Z_N}\int \frac{d\mu}{\sqrt{4\pi/\beta\phi N}}e^{-\frac{\beta\phi N}{4}\mu^2}\int\frac{d\xi}{2\pi/N} \sum_{\{\bm\sigma,\bm\tau\}} \prod_i \left(e^{ F_{\sigma_i}(m_\sigma)\tau_i}e^{\I\xi\sigma_i}\right)e^{-\I\xi m_\sigma N}\int \frac{d\zeta}{2\pi/N} e^{-\I\zeta (Nm_\tau-\sum_i\tau_i)}\nonumber\\
&= \frac{1}{Z_N}\int \frac{d\mu}{\sqrt{4\pi/\beta\phi N}}e^{-\frac{\beta\phi N}{4}\mu^2}\int\frac{d\xi d\zeta}{(2\pi/N)^2} \sum_{\{\bm\tau\}} \prod_i \left[\left(e^{ F_{+}(m_\sigma)\tau_i} e^{\I\xi}+e^{F_{-}(m_\sigma)\tau_i} e^{-\I\xi}\right)e^{\I\zeta \tau_i}\right]e^{-\I N(\xi m_\sigma +\zeta m_\tau)}\nonumber\\
&= \frac{1}{Z_N}\int \frac{d\mu}{\sqrt{4\pi/\beta\phi N}}e^{-\frac{\beta\phi N}{4}\mu^2}\int\frac{d\xi d\zeta}{(2\pi/N)^2}  \left[\left(e^{ F_{+}(m_\sigma)} e^{\I\xi}+e^{F_{-}(m_\sigma)} e^{-\I\xi}\right)e^{\I\zeta} + 2 \cos\xi + \right. \nonumber\\
&\quad  \left. +\left(e^{-F_{+}(m_\sigma)} e^{\I\xi}+e^{-F_{-}(m_\sigma)} e^{-\I\xi}\right)e^{-\I\zeta}\right]^N e^{-\I N(\xi m_\sigma +\zeta m_\tau)}\ .
\end{align}

\end{document}